\documentclass[10pt]{iopart}
\pdfoutput=1

\usepackage{iopams}  
\usepackage[hypertexnames=false, unicode, colorlinks, allcolors=blue!70!black, linktocpage, pdfusetitle]{hyperref}
\usepackage{amssymb,amsbsy}
\usepackage{bm}
\usepackage{soul}
\usepackage{color,xcolor}
\usepackage[pdftex]{graphicx}
\usepackage[numbers,square,comma,sort&compress]{natbib}
\usepackage{url}
\usepackage[T1]{fontenc}
\usepackage[utf8]{inputenc}
\usepackage{lmodern}

\newcommand{\be}{\begin{equation}}
\newcommand{\ee}{\end{equation}}
\newcommand{\bea}{\begin{eqnarray}}
\newcommand{\eea}{\end{eqnarray}}
\newcommand{\onehalf}{\textstyle{\frac{1}{2}}}

\newcommand{\lp}{\left (}
\newcommand{\rp}{\right)}

\newcommand{\m}{-}
\newcommand{\p}{+}

\begin{document}

\title[Numerical Relativity with Arbitrary Precision Arithmetic]{Numerical Relativity with Arbitrary Precision Arithmetic: Applications to Gravitational Collapse}

\author{Daniel Santos-Oliv\'an, Carlos F.~Sopuerta}

\address{Institut de Ci\`encies de l'Espai (ICE, CSIC), Campus UAB, Carrer de Can Magrans s/n, 08193 Cerdanyola del Vall\`es, Spain}
\address{Institut d'Estudis Espacials de Catalunya (IEEC), Edifici Nexus I, Carrer del Gran Capit\`a 2-4, despatx 201, 08034 Barcelona, Spain}
\ead{dsantosolivan@gmail.com,sopuerta@ice.csic.es}

\vspace{10pt}
\begin{indented}
\item[]\today
\end{indented}

\begin{abstract}
Numerical Relativity is a mature field with many applications in Astrophysics, Cosmology and even in Fundamental Physics. 
As such, we are entering a stage in which new sophisticated methods adapted to open problems are being developed. 
In this paper, we advocate the use of Pseudo-Spectral Collocation (PSC) methods in combination with high-order precision arithmetic for  Numerical Relativity problems with high accuracy and performance requirements. 
The PSC method provides exponential convergence (for smooth problems, as is the case in many problems in Numerical Relativity) and we can use different bit precision without the need of changing the structure of the numerical algorithms. Moreover, the PSC method provides high-compression storage of the information.  
We introduce a series of techniques for combining these tools and show their potential in two problems in relativistic gravitational collapse: (i) The classical Choptuik collapse, estimating with arbitrary precision the location of the apparent horizon. (ii) Collapse in asympotically anti-de Sitter spacetimes, showing that the total energy is preserved by the numerical evolution to a very high degree of precision.
\end{abstract}

\vspace{2pc}
\noindent{\it Keywords}: Numerical Relativity, Gravitational Collapse, Pseudospectral Collocation Methods, Arbitrary-precision arithmetic
%
\vspace{10pt}

\submitto{\CQG}
%
%
%

\section{Introduction}
\label{introduction}

Numerical Relativity emerged as a way to provide answers to problems in General Relativity that require solving Einstein equations in physical situations where we do not have neither exact solutions nor good approximation schemes, or at least not at the level of precision that is required. In particular, for systems characterized either by extreme gravitational fields or by relativistic speeds, or both.  After more than five decades of developments, Numerical Relativity has become a mature field that has produced revolutionary discoveries for very different types of problems (for some accounts of Numerical Relativity can found in~\cite{Lehner:2001wq,Grandclement:2007sb,Alcubierre:1138167,Bona:2009bo,Baumgarte:2010bs,Lehner:2014asa,Cardoso:2014uka}: From gravitational collapse to cosmological physics, including the description of the non-linear dynamics of binary systems and their gravitational wave emission. Clearly, the problems that Numerical Relativity has addressed are very demanding in terms of the complexity of the algorithms (including formulating a well-posed initial-value problem) and also in terms of computational cost. This has lead to the development of more and more sophisticated algorithms that use state-of-the-art techniques from the fields of numerical analysis and computer science. Nevertheless, there is a question that has not been sufficiently discussed in Numerical Relativity, the question of numerical accuracy in relation to the numerical precision offered by current digital computers. The modelling of physical phenomena requires choosing the appropriate type of numerical algorithm that ensures a satisfactory answer in terms of accuracy, reliability, and computational cost.  For many problems, the common $64$-bit (double precision) floating-point arithmetic, i.e. fifteen or sixteen significant digits, is enough to obtain accurate results. In some cases, even $32$-bit (single precision) floating-point arithmetic, i.e. seven or eight significant figures, can be sufficient.  However, there are problems that demand a very high degree of precision~\cite{math3020337}.  For instance, in cases where solutions at late times strongly depend on the initial conditions or in cases where the physical properties are very sensitive to the value of certain parameters, the ability to increase the precision of the numerics can be an essential ingredient to reach satisfactory results (see, e.g.~\cite{Khanna:2013}).

Different types of numerical algorithms have been used in numerical relativity. Finite Differences provide simple and easy ways to design algorithms while Finite Element methods are in general more robust and modular, which is the reason why they are very frequently used in engineering problems. Nevertheless, if the priority is to achieve high accuracy, then spectral and pseudospectral methods are a convenient choice due to their great convergence properties: They converge exponentially for smooth problems. It has also been shown that they provide highly precise solutions in a variety of problems, from fluid dynamics to astrophysics (see, e.g.~\cite{canuto2007spectral,Bourke1988,ehrendorfer2012spectral}). In Numerical Relativity, they have successfully been applied to the simulation of the collision of orbiting binary black holes and their emission of gravitational waves~\cite{Szilagyi:2009qz,Haas:2016cop} (see also e.g.~\cite{Grandclement:2007sb,Canizares:2009ay,Canizares:2010yx,Macedo:2014bfa,Oltean:2018szc}). 

The main goal of this paper is to show that the powerful convergence properties of PseudoSpectral Collocation (PSC) methods make them an ideal option to go beyond the typical $64$-bit floating-point arithmetic for problems in the context of Numerical Relativity. 
For the common {\em double} precision, maximum accuracy is usually reached with a relatively quite low number of collocation (discretization) points.  Therefore, going beyond this precision does not constitute a high increase in computational cost. In addition, PSC methods provide a high-compression of the information describing the solution of our problem. Indeed, the number of collocation points (or modes in the spectral picture) needed to reach very high accuracy is much smaller than the number of points in other methods (e.g., Finite Differences) so that the memory demands get reduced drastically.  Moreover, the structure of the algorithms, and in consequence of the numerical codes that implement them, is independent of the number of collocation points, which is also another advantage of the PSC method since we do not have to touch the algorithms in order to increase the precision, just to change the number of collocation points.  This is contrast with Finite Differences methods, where by increasing the number of discretization points, within a wide range, we reach machine precision and then a new Finite Differences algorithm with better truncation error would have to be implemented (i.e. with a different stencil that would require, in general, more discretization points).

In order to show in practice all these claims, we have developed a new numerical library, ANETO (Arbitrary precisioN solvEr with pseudo-specTral MethOds)~\cite{anetolib}, that we use to perform the numerical simulations described in this paper. The ANETO library provides complete freedom in the numerical precision in the sense that we have the possibility of adjusting the bit precision of our algorithms to fulfil the exact accuracy requirements of a given problem. It also contains a tool to translate numerical codes from the standard double precision to arbitrary precision. The work presented in this paper originates from the experience of the authors in the study of gravitational collapse in several scenarios that require the use of General Relativity (see~\cite{Olivan:2015fmy,SantosOlivan:2016djn}), and where the accuracy requirements on the numerical solutions are very high.  In the sense, we have developed numerical codes that use the ANETO library to study two different scenarios of gravitational collapse: 
(i) The collapse of a spherically-symmetric massless scalar field in asymptotically-flat spacetimes, the scenario where Choptuik found critical behaviour~\cite{Choptuik:1992jv}. We have developed a new characteristic pseduospectral code using the ANETO library and show that we can essentially have arbitrary precision in the estimation of the location where an apparent horizon is formed.
(ii) The collapse of a spherically-symmetric massless scalar field in asymptotically-Anti de Sitter (AdS) spacetimes, which have recently attracted a lot fo attention in the context of string theory and the gauge/gravity duality. The fact that AdS, the maximally symmetric solution of Einstein's equations with a negative cosmological constant (see more in~\cite{Hawking:1973uf}), has the remarkable property that light rays can reach the AdS boundary in a finite time (in contrast to massive particles that need an infinite time as in flat spacetime)    has important consequences for the study of gravitational collapse. At the same time, these numerical studies demand a very high degree of precision. We adapt the code used in~\cite{Olivan:2015fmy,SantosOlivan:2016djn} to be compatible with the ANETO library to study the long-term dynamics in AdS spacetimes, showing that we can keep the spacetime  mass constant to very high degrees of precision.

The plan of this paper is as follows: 
In Sec.~\ref{sec_basics_PSC} we introduce the main techniques that we have developed to combine the PSC method with arbitrary precision arithmetic for the study of the dynamics of the general relativistic gravitational collapse. This includes the main operators, a multidomain scheme, a discussion on computational time and parallelization. 
In Sec.~\ref{grav-collapse-Minkowski}, we report on simulations of gravitational collapse in asymptotically-flat spacetimes using the ANETO library and we do the same in Sec.~\ref{grav-collapse-anti-de-Sitter} for the case of gravitational collapse in asymptotically-anti de Sitter spacetimes.
We finish with some conclusions and a discussion of future prospects in Sec.~\ref{conclusions-perspectives}.

\section{The Numerical Method: PseudoSpectral Collocation Methods with Arbitrary Precision Arithmetic}\label{sec_basics_PSC}

There are problems in the context of NR that require high precision numerical computations, in which the usual {\em double precision} used in most codes is not enough.  To deal with this type of demanding problems we have to consider the possibility of using other representations of real numbers with a larger number of significant digits, taking always into account the limited memory of a digital computer. The usual approach is based on the so-called {\em floating-point} representation (see~\cite{ieee745-2019}), similar to scientific notation, where a real number $x$ can be expressed/approximated by:
\begin{equation}
fl(x) = \left( 1 + \sum_{i=1}^{b^{}_m-1} b^{}_i\;\times 2^{-i} \right) \;\times\; 2^E\,,
\label{formula_fpa}
\end{equation}
where $i$ is the position of the bit $b^{}_{i}$ of the mantissa from the left, $b_m$ is the number of bits of the mantissa (the precision), and $E$ is the number of bits of the exponent. Then, $E$ establishes the maximum range of our variables and $b_m$ determines the machine precision (or machine roundoff error), which is what we are interested in. A measure of the level of roundoff error in the floating-point number system is:
\begin{equation}
\epsilon_{\rm mp}= \max_{x \neq 0} \frac{|x-fl(x)|}{|x|} \,,
\end{equation}
When rounding is made by chopping, we have that $\epsilon_{\rm mp}=2^{1-b_{m}}$. In this work, the standard numerical precisions that we consider for reference are: {\em single precision}, where $b_{m} = 24$ (eight significant digits); {\em double precision}, with $b_{m} = 53$, corresponding to $15-16$ significant digits; and {\em quadruple precision}, where $b_{m} = 113$, corresponding to approximately $34$ significant digits. Apart from that, we also use arbitrary precision so that we manually select the bit precision $b_{m}\,$.

On the other hand, given the additional computational cost of high-order arithmetic, we have to think carefully what type of discretization algorithms we use to solve the partial differential equations involved in our physical problem.  There two important factors, one is the question of the computational cost of the operations and the size of the storage required to store the information associated with the variables involved in the computation. The other one is the question of whether we need to adapt the discretization algorithms (and their programming language implementation) in terms of the particular numerical precision chosen. In this paper we advocate for the use of the PSC method~\cite{Boyd,Fornberg:1996psc,Canutoetal:2006sm1,Grandclement:2007sb} for spatial discretization of our variables. The reason is that due to spectral convergence of the PSC method for smooth problems we just need in general much less grid points as compared with other discretization methods, which reduces the storage needs.  On the other hand, we do not need to change the  algorithm code in the PSC method to increase the precision, just the number of collocation points.  All the PSC algorithms that use arbitrary precision arithmetic in this work have been developed using the ANETO library. The library has been developed using C++ templates, which allows for the use of any kind of data type.  In several of the computations done for this paper, we have used standard types like {\em float} and {\em double} for single and double precision respectively. For the {\em quadruple precision} type we have used the {\em float128} type implemented in the Boost Multiprecision library~\cite{boostmultiprecisionlib}. For general bit precision, we have used the GNU's Multiple Precision Floating-Point Reliable (MPFR) Library~\cite{Fousse:2007}, a C library for multiple-precision floating-point computations with correct rounding, with a C++ wrapper~\cite{mpfrcpp} as an interface. Some numerical algorithms like differentiation and integration has been paralellized using the shared-memory API OpenMP~\cite{dagum1998openmp}.
Finally, the ANETO library has been released as Free Software under a GNU General Public License (GPL) and it can be found in~\cite{anetolib}, including source code and full documentation of the classes and functions available as well as examples of its functionalities.  The library was originally developed due to need to go beyond the standard double precision for the study of certain problems associated with gravitational collapse in the context of General Relativity.

\begin{figure}[t]
\begin{center}
\resizebox{.6\textwidth}{!}
{ 
\includegraphics{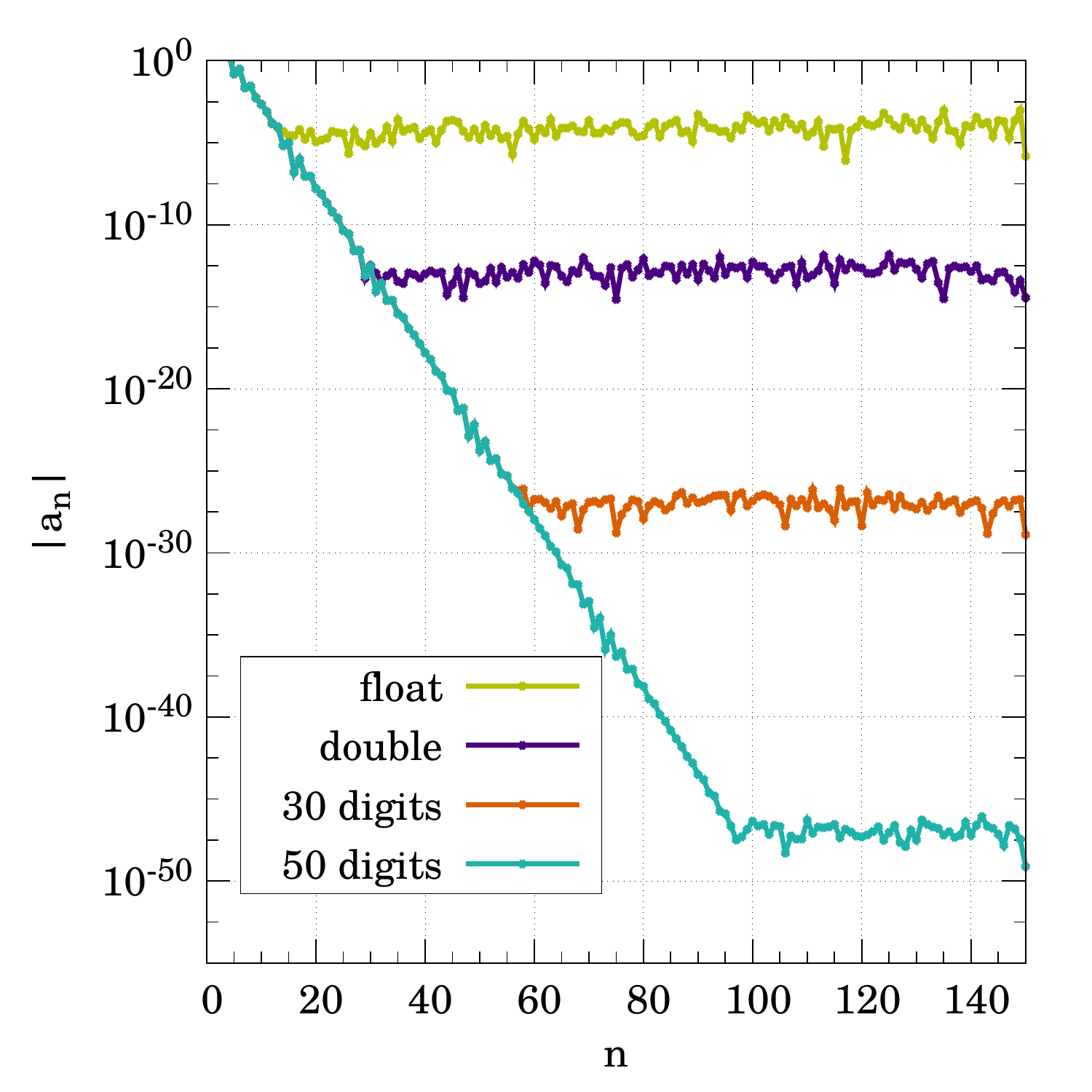}
}
\caption[Representation of the Spectral Coefficients]{
{\bf Representation of the Spectral Coefficients.} For a smooth function (we have used the function $f(x) = \exp(\tan(x))$) the modes in the spectral representation decay exponentially until they reach the precision error of our computations. We can see that with the use of arbitrary precision arithmetic we can control the level of the computer round-off error.\label{fig_spec_coeffs}}
\end{center}
\end{figure}

The type of problems we are interested in constitute an initial-boundary value problem, which consists in a system of Partial Differential Equations (PDEs) defined over a spatial domain ${\cal D} \in \mathbb{R}^{d}$, being $d$ the number of space dimensions, and for a time interval ${\cal T}\equiv [t_{i},t_{f}]\subset\mathbb{R}$:
\begin{eqnarray}
{\cal L}[\vec{u}](t,\vec{x}) = 0\,, 
\quad
{\cal I}[\vec{u}](t_{i},\vec{x}) = 0\,, 
\quad
{\cal B}[\vec{u}](t,\vec{W}(t,\vec{x})) = 0\,, \label{efes-general}
\end{eqnarray}
where $t\in{\cal T}\,,\, \vec{x}\in{\cal D}$, and where $u(t,\vec{x})$ denotes the vector of unknown variables; ${\cal L}$ is a given differential operator that determines the set of PDEs under consideration; ${\cal I}$ is another given operator representing the initial conditions of our evolution problem at $t=t_{i}$; and finally, ${\cal B}$ is the operator that determines the boundary conditions at a set of (timelike) hypersurfaces defined by a set of implicit equations $\vec{W}(t,\vec{x})=\vec{0}$. The operator ${\cal L}$ can be divided into two independent parts: One producing a set of hyperbolic evolution equations and the other hand producing a set of constraint equations that the initial data (expressed here in terms of the operator ${\cal I}$) has to satisfy and the evolution has to preserve.
In spectral methods the solution to this type of problems is approximated by using a spectral expansion of the form:
\begin{equation}
\vec{u}^{}_{N}(t,x) = \sum_{k=0}^{N} \vec{a}^{}_{k}(t)\, \phi^{}_k(x)\,,  \label{chap_approx_spectral_repr}
\end{equation} 
where $\phi_k$ are the basis functions and $\vec{a}_k$ are the vectors of (spectral) coefficients associated with the approximation to the vector of unknowns, $\vec{u}_{N}(t,x)$. 
In this paper we choose the basis functions $\phi^{}_k$ to be Chebyshev polynomials: 
\begin{equation}
T^{}_n(X) =\cos\left[n\cos^{-1}(X) \right] \quad (n=0,\ldots\,N; X \in [-1,1])\,.
\label{chap_chebyshev-polynomials}
\end{equation}
Notice that here, the coordinate $X$, which we call the {\em spectral} coordinate, will not be in general the same as the coordinate $x$ in Eq.~(\ref{chap_approx_spectral_repr}), which we call the {\em physical} coordinate.  They will have in general different ranges and will be related by a one-to-one mapping\footnote{Although we are not considering it here, it is in principle possible to include a dependence on the time $t$ in this mapping.}, $x=x(X) \Leftrightarrow X=X(x)\,$. 
The PSC method consists in finding the solution by demanding that our equations~(\ref{efes-general}) are exactly satisfied at a set of collocations points. In this work we consider only 
the {\em Lobatto-Chebyshev} grid of collocation points:
\begin{eqnarray}
X^{}_i = -\cos\left(\frac{\pi\,i}{N}\right) \quad (i=0,\ldots,N)\,,
\label{chebyshevlobattogrid}
\end{eqnarray}
This particular choice, apart from minimizing the interpolation error, includes the boundary points, $X=\pm 1$, which allows us to directly impose boundary conditions there.
In the PSC method we have two representations of the approximation to the solution of our problem.  The first one, given by Eq.~(\ref{chap_approx_spectral_repr}), is the standard one in spectral methods and hence we call it the {\em spectral} representation. The magnitude of the spectral coefficients decays exponentially with the degree, $n$, of the Chebyshev polynomial:
\begin{equation}
|a^{k}_n| \sim e^{- \alpha^{k}\, n} \quad (n=0,\ldots,N)\,,
\end{equation}
where $\alpha^{k}$ is a coefficient associated with the variable $u^{k}$. As a consequence, the discretization error due to the use of a finite number of points also decays exponentially. This behaviour is illustrated in Figure~\ref{fig_spec_coeffs} for numerical computations using different number of digits of precision.  

The other type of representation is the {\em physical} representation, based on the use of the collocation values of our variables, $\vec{u}^{}_{i}$, and given by: 
\begin{equation}
\vec{u}^{}_{N}(t,x) = \sum_{i=0}^N \vec{u}^{}_{i}(t)\, {\cal C}^{}_i(X)\,, 
\label{physicalrepresentation}
\end{equation}
where $\vec{u}^{}_{i}(t)$ are the values of our variables $\vec{u}$ at the collocation point $X^{}_{i}$, and ${\cal C}^{}_i(X)$ are the {\em cardinal} functions~\cite{Boyd} associated with our basis functions and grid, obeying: ${\cal C}^{}_{i}(X^{}_{j}) = \delta^{}_{ij}$. 
The two representation are related via a matrix transformation:  
\begin{equation}
\vec{u}^{}_{i} = \sum_{n=0}^N \mathbb{M}^{}_{in} \vec{a}^{}_{n}\,, 
\quad  
\mathbb{M}^{}_{in} = T^{}_n(X^{}_{i}) \quad (i,n=0,\ldots,N) \,.
\label{matrix-transformation}
\end{equation}
Introducing a new coordinate via $X = \cos\theta$ ($\theta\in[0,\pi]$), the series in Chebyshev polynomials [Eq.~(\ref{chap_chebyshev-polynomials})] becomes a cosine series since $T^{}_n(\cos\theta) =\cos\left(n\theta \right)$. In the particular case of the Lobatto-Chebyshev grid [Eq.~(\ref{chebyshevlobattogrid})], the components of the matrix transformation are: $\mathbb{M}^{}_{in} = T^{}_n(X^{}_{i})= (-1)^{n}\cos(n\,i\,\pi/N)$. The resulting transformation becomes then a Discrete Cosine Transform (DCT) that can be computed by using a $2N$ Fast Fourier Transform (FFT) algorithm whose computation cost scales as $\sim N\ln N$ with the number of collocation points~\cite{canuto1988spectral,Boyd} in contrast with the $\sim N^{2}$ scaling of the matrix transformation. 
The possibility of using the FFT algorithm to change between the physical and spectral representations is of particular relevance for the solution of time-dependent problems described by PDEs.  The physical representation is useful to deal with non-linear terms since the collocation values associated with an arbitrary function of $\vec{u}$, say $\vec{f}(\vec{u})$, are simply given by $\vec{f}(\vec{u})^{}_{i} = \vec{f}(\vec{u}^{}_{i})$. In contrast, the spectral representation is better for other type of operators like differentiation and integration.

\subsection{Differentiation and Integration}\label{subsec_PSC_diff}
In the PSC method, differentiation can performed by means of the following schematic procedure:
\begin{eqnarray}
\partial^{}_{x} :\,\{\vec{u}^{}_i\} ~\stackrel{\rm DCT}{\longrightarrow}~
\{\vec{a}^{}_n\} ~\stackrel{\tilde\partial^{}_{x}}{\longrightarrow}~
\{\vec{b}^{}_n\} ~\stackrel{\rm DCT}{\longrightarrow}~
\{(\partial^{}_{x}{\vec{u}})^{}_i\}\,, \label{pscdifferentiation}
\end{eqnarray}
where $\tilde\partial^{}_{x}$ is the spectral-space derivative operator whose action is given by~\cite{Boyd,canuto1988spectral}
\begin{equation}
\left\{
\begin{array}{lcl}
\vec{b}^{}_{N}    & = & 0\,, \\[1mm]
\vec{b}^{}_{N\m1} & = & 2 N \vec{a}^{}_{N}\,, \\[1mm]
\vec{b}^{}_{n}    & = & \frac{1}{\bar{c}^{}_{n}}\left[ 2(n+1)\,\vec{a}^{}_{n+1}+\vec{b}^{}_{n+2}  \right] \quad (n = N\m2\ldots 0)\,,
\end{array}
\right.
\label{spectralderivativeoperator}
\end{equation}
where the $\bar{c}^{}_{n}$ are such that: $\bar{c}^{}_{n}=2$ if $n=0$ or $n=N$ and $\bar{c}^{}_{n}=1$ otherwise.

In the case of integration, there are different relevant operators. First, let us consider the following simple first-order ODE
\begin{equation}
\frac{d f(x)}{d x} = g(x)\,,
\end{equation}
with a boundary condition at $x=x_{0}$, $f(x_{0}) = f_{0}$.  The solution is simply given by
\begin{equation}
 f(x) = f_{0} + \int_{x_0}^{x} dx'\,g(x')  \,.
\end{equation}
This is what we call {\em integration from the left} because it incorporates a (boundary) condition at $X = X^{}_{\m} = -1$ (assuming a mapping $x=x(X)$ so that $x_{0}=x(X^{}_{\m})$):
\begin{equation}
I^{}_{\rm L}(x(X)) = I^{}_{\rm L}(x^{}_{0}) + \int^{X}_{X^{}_{\m}}dX'\,\left(\frac{dx}{dX}\right)^{}_{X'}g(X') \,,
\label{integr_IL}
\end{equation}
where $I^{}_{\rm L}(x_{0})=f_{0}$.  The scheme for the full left-integration process is:
\begin{eqnarray}
\begingroup\textstyle\int\endgroup^{X}_{X_{\m}} :\,\{\vec{u}^{}_i\} ~\stackrel{\rm DCT}{\longrightarrow}~
\{\vec{a}^{}_n\} ~\stackrel{\int^{}_{\rm L}}{\longrightarrow}~
\{\vec{b}^{\rm L}_n\} ~\stackrel{\rm DCT}{\longrightarrow}~
\{(\int^{X}_{X_{\m}}{\vec{u}})^{}_i\}\,, 
\label{pscintegration-left}
\end{eqnarray}
where $\{\vec{b}^{\rm L}_{n}\}$ are the spectral coefficients corresponding to $I^{}_{\rm L}(x(X))$:
\begin{equation}
\left\{
\begin{array}{lcl}
\vec{b}^{\rm L}_{N} & = & \frac{\vec{a}^{}_{N\m1}}{2N}\,, \\[1mm]
\vec{b}^{\rm L}_{n} & = & \frac{1}{2n}\left(\bar{c}^{}_{n-1}\,\vec{a}^{}_{n\m1} - \vec{a}^{}_{n+1}\right) \quad \left(n = N-1,\ldots,1\right)\,,  \\[1mm]
\vec{b}^{\rm L}_{0} & = &  I(x(X^{}_{\m})) - \sum^{N}_{n=1} \lp -1\rp^n \vec{b}^{}_{n} \,.
\end{array}
\right.
\label{intspec2}
\end{equation}
In the same way, we can introduce the right integration (the boundary condition is imposed at the right boundary, i.e. at $X = X^{}_+ = +1$):
\begin{equation}
I^{}_{\rm R}(x(X)) = I^{}_{\rm R}(x^{}_{0}) + \int^{X^{}_{+}}_{X}dX'\,\left(\frac{dx}{dX}\right)^{}_{X'}g(X')\,.
\label{integr_IR}
\end{equation}
The scheme for the full right-integration process is:
\begin{eqnarray}
\begingroup\textstyle\int\endgroup^{X^{}_{+}}_{X} :\,\{\vec{u}^{}_i\} ~\stackrel{\rm DCT}{\longrightarrow}~
\{\vec{a}^{}_n\} ~\stackrel{\int^{}_{\rm R}}{\longrightarrow}~
\{\vec{b}^{\rm R}_n\} ~\stackrel{\rm DCT}{\longrightarrow}~
\{(\int^{X^{}_{+}}_{X}{\vec{u}})^{}_i\}\,.
\label{pscintegration-right}
\end{eqnarray}
The action of the operator $\int^{}_{\rm R}$ is given by
\begin{equation}
\left\{
\begin{array}{lcl}
\vec{b}^{\rm R}_{N} &=& - \frac{1}{2 N} \vec{a}^{}_{N\m1}, \\[1mm]
\vec{b}^{\rm R}_{n} &=& - \frac{1}{2 n} \lp c^{}_{n\m1} \, \vec{a}^{}_{n\m1}  - \vec{a}^{}_{n\p1}\rp \quad \left(n = N-1,\ldots,1\right) \,,
\\[1mm]
\vec{b}^{\rm R}_{0} &=&  I^{}_{\rm R}(x(X^{}_{+})) - \sum_{n=1}^{N} \vec{b}^{\rm R}_n \,.
\end{array}
\right.
\end{equation}
%

\subsection{The Multidomain PSC Method} \label{subsection_multidom}

In many problems, different regions of our computational domain require different degrees of spatial resolution and hence some form of grid refinement needs to be adopted. In the case of the PSC method, a simple choice is to use a multidomain scheme that distributes the subdomains so that regions where high resolution is required are covered by more subdomains than regions less demanding in terms of resolution.  For the case of evolution problems in one spatial dimension we consider a decomposition of our physical computation domain $\Omega=[x^{}_{L},x^{}_{R}]$ in $D$ disjoint subdomains:
\begin{equation}
\Omega = \bigcup^{D-1}_{a=0} \Omega^{}_a\,, ~~\Omega^{}_a = \left[ x^{}_{a,L}, x^{}_{a,R}\right]\,,
\end{equation}
with the identification: $x^{}_{a,R}=x^{}_{a+1,L}$ ($a=0,\ldots,D-2$). Each subdomain $\Omega^{}_a$ is mapped into the spectral domain $[-1,1]$. For our computations we assume a simple linear mapping but other mappings are possible. Then, given the coordinate $x$ of a point in the physical domain $\Omega$, assuming it belongs to the subdomain $\Omega^{}_a$, it is mapped to a spectral coordinate $X^{}_{a}$ according to the following linear mapping:
\begin{eqnarray}
x \longrightarrow X^{}_a(x) = \frac{2x- x^{}_{a,L}- x^{}_{a,R}}{ x^{}_{a,R} - x^{}_{a,L} } \,.    \label{lmap1}
\end{eqnarray} 
whose inverse mapping is:
\begin{eqnarray}
X^{}_a  \longrightarrow x(X^{}_a) = \frac{x^{}_{a,R}-x^{}_{a,L}}{2}X^{}_a + \frac{x^{}_{a,L}+ x^{}_{a,R}}{2}\,. \label{lmap2}
\end{eqnarray}
The Jacobian, ${d x}/{d X^{}_a}  = (x^{}_{a,R}-x^{}_{a,L})/{2}$, is different from subdomain to subdomain unless all of them have the same physical coordinate size.  Despite its simplicity, the linear mapping can be used for refinement adapting the length of each subdomain to the resolution needs of the problem under consideration.  All what we need is a refinement criteria for the adaptation of the subdomain sizes.

The important point of using a multidomain scheme for solving PDEs is communication.  In the case of elliptic equations we can just communicate variables by imposing continuity conditions. In the case of hyperbolic PDEs, assuming they are strongly hyperbolic~\cite{Courant:1989aa,John:1991fj}, which means the principal part operator has a complete set of eigenvectors and all the eigenvalues, the propagation speeds of the eigenvalue fields (known as characteristic fields) are real. Then, at the interface between two subdomains, the fields with positive speed are communicated from the left to the right subdomain, while the fields with negative speeds are communicated from the right to the left subdomain. This scheme can even accommodate discontinuities like those produced by the presence of point particles (see~\cite{Sopuerta:2005gz,Canizares:2010yx,Canizares:2011kw,Oltean:2018szc}). Finally, the characteristic variables are can also be used to impose boundary conditions at the global boundaries in a clear and simple manner, for instance for the case of in/outgoing boundary conditions.

\begin{figure}[t]
\begin{center}
\resizebox{.6\textwidth}{!}
{
\includegraphics{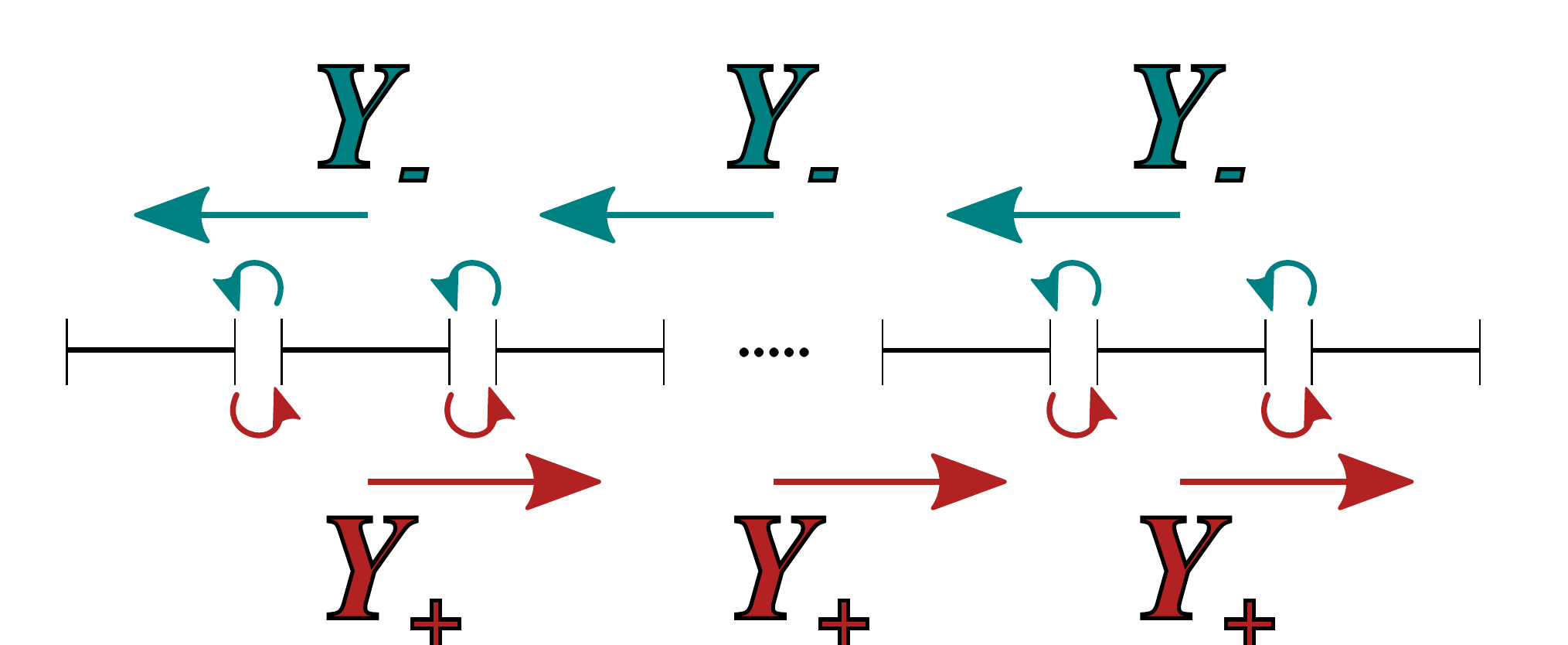}
}
\caption[Communication between subdomains]{\textbf{Communication between subdomains.} Schematic representation of the communication between subdomains for hyperbolic PDEs. Characteristic variables, $Y_-$ and $Y_+$, with a well-defined speed of propagation are the key. We just need to copy the boundary values in the direction indicated by the arrows. \label{plot_diagram_U_V}}
\end{center}
\end{figure}

\begin{figure}[t]
\begin{center}
\resizebox{0.9\textwidth}{!}
{ 
\includegraphics{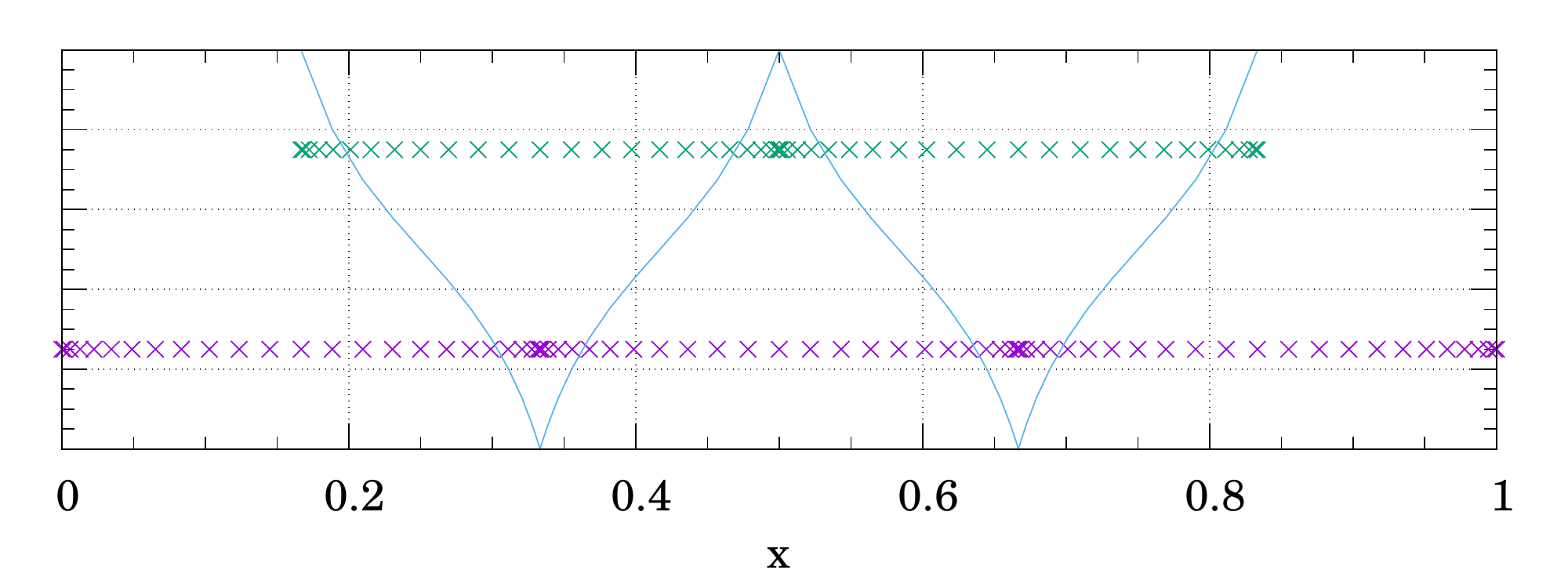}
}
\caption[Dual Grid Structure]{{\bf Dual Grid Structure.} Schematic representation of the dual grid scheme (without showing the subdomains that have the global boundaries). The main grid (in purple) is composed of three subdomains and the dual grid (in green) is shifted and has four subdomains. The blue line shows the weighting function $\pi^{}_{a}(x)$ of Eq.~(\ref{partition}). \label{fig_dualgrid_scheme}
}
\end{center}
\end{figure}

Although many operations generalize trivially to a PSC multidomain scheme in the sense that they can be performed independently at each subdomain, some others require some adaptation that has to do again with communication between subdomains. Differentiation is a good example. It is well-known that the computation of derivatives with the PSC method becomes noisy near the boundaries producing an accumulation of error there, which in some cases can be one or two orders of magnitude higher than in other more central regions.  This phenomenon, known as Runge's phenomenon~\cite{Runge:1901cfr,Epperson:1987je}, gets worse as we increase the number of collocation points.  We can take advantage of the multidomain scheme to reduce the error by using a dual grid scheme. The idea is to use a second grid constructed such that the boundaries of the subdomains coincide with the middle points of the subdomains of the main (original) grid (see Figure~\ref{fig_dualgrid_scheme}). As a consequence the dual grid has $D+1$ subdomains, $\left\{\bar{\Omega}^{}_{\bar{a}}\right\}^{}_{\bar{a}=0,\ldots,D}$, with the left boundary location of each subdomain, say $\bar{a}$ ($=0,\ldots, D$), being given by $\bar{x}^{}_{\bar{a},L} = (x^{}_{\bar{a}-1,L} + x^{}_{\bar{a}-1,R})/2$ (excepting for $\bar{a}=0$, which corresponds to the global left boundary point $x^{}_{L}$), and the right boundary location is given by $\bar{x}^{}_{\bar{a},R} = (x^{}_{\bar{a},L} + x^{}_{\bar{a},R})/2$ (excepting for $\bar{a}=D$,  which corresponds to the global right boundary point $x^{}_{R}$). When we compute the derivative in the dual grid, the points where typically the error is the lowest coincide with the location where the error is typically the greatest in the subdomains of the main grid.  Then, we compute the final derivative of our function by combining the derivatives computed separately in the main ($f'^{}_{\rm main}$) and dual ($f'_{\rm dual}$) grids.  For a given point $x$ belonging to the subdomain $\Omega^{}_{a}$ of the main grid, the derivative is the following weighted sum of the derivatives from the main and the dual grids:
\begin{equation}
f'(x)  =   \pi^{}_{a}(x) f'^{}_{\rm main}(x) + \left(1-\pi^{}_{a}(x)\right) f'_{\rm dual}(x)
\quad 
(x \in \Omega^{}_{a})\,, 
\label{derivative_dual} 
\end{equation}
where $\pi^{}_{a}(x)$ is a weighting function on the subdomain $\Omega^{}_{a}$ of the main grid (this function together with $1-\pi^{}_{a}(x)$ form a partition of unit associated with the subdomain $\Omega^{}_{a}$) that takes values between zero and one, being zero in the boundaries of the subdomains of the main grid and one at the boundaries of the dual grid. Moreover, it is smooth between the boundary and the centre of the subdomain. One example of a partition function is:
\begin{equation}
\pi^{}_{a}(x) = \left\{ 
\begin{array}{ll}
\frac{(x - x^{}_{a,L}) (x - x^{}_{a,R})}{(x - x^{}_{a,L})  (x - x^{}_{a,R}) + (x - \bar{x}^{}_{a,L}) (x - \bar{x}^{}_{a,R})}  
& \mbox{if}~~x~\in~\bar{\Omega}^{}_{a} \,,\\[3mm]
\frac{(x - x^{}_{a,L}) (x - x^{}_{a,R})}{(x - x^{}_{a,L})  (x - x^{}_{a,R}) + (x - \bar{x}^{}_{a+1,L}) (x - \bar{x}^{}_{a+1,R})}  
& \mbox{if}~~x~\in~\bar{\Omega}^{}_{a+1}\,.
\end{array}
\right.
\label{partition}
\end{equation}
This structure of $\pi^{}_{a}(x)$ is due to the fact that a given point of the subdomain $\Omega^{}_{a}$ can either be at the subdomain $\bar{\Omega}^{}_{a}$ or $\bar{\Omega}^{}_{a+1}$ of the dual grid (see Figure~\ref{fig_dualgrid_scheme}).

\begin{figure}[t]
\begin{center}
\resizebox{1\textwidth}{!}
{ 
\includegraphics{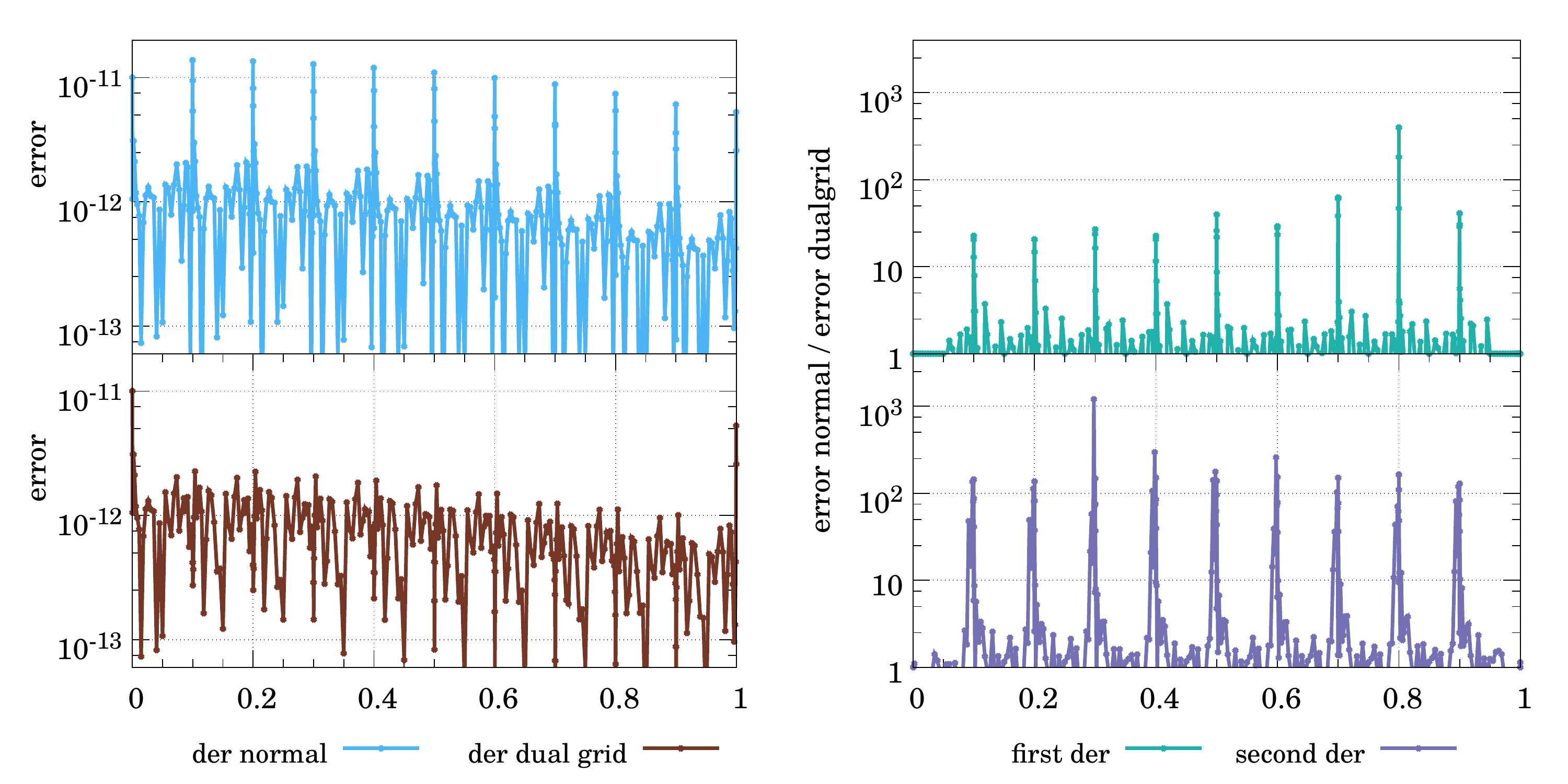}
}
\caption[Differentiation using a Dual Grid]{\textbf{Differentiation using a Dual Grid.} The plots on the left show the error in the first-order derivative computed with the dual-grid setup (bottom) and without it (top).  We see that at the boundaries across subdomains there is an efficient reduction of the error by eliminating the peaks. The magnitude of the improvement can be seen in the top right plot. The quotient of the error in both cases tells us that the improvement at the points near the boundaries is between one and two orders of magnitude. The bottom right plot shows the same for the second-order derivative where the improvement increases by two or three orders of magnitude.}
\label{fig_dualgrid_error}
\end{center}
\end{figure}

We have carried out several numerical experiments to assess the use of the dual-grid structure by comparing with computations on a single grid.  The results are summarized in Figure~\ref{fig_dualgrid_error}.  In the top-left plot we show the error in the computation of a derivative using ten subdomains. It is easy to find in the plot the location of the boundaries between the subdomains since the error exhibits there a peak that in most cases is one order of magnitude higher than the average. The brown line of the bottom-left plot shows the error when we perform the same computation but using the dual-grid scheme. As we can see, the peaks have completely disappeared and the error in the derivative looks now quite flat. This improvement can be seen on the top-right plot of Figure~\ref{fig_dualgrid_error}, where we present the ratio between the differentiation errors with and without using the dual grid. Near the boundaries our computation has improved between one and two orders of magnitude in terms of the error. We have done the same comparison for the second-order derivative, which is shown in the bottom-right plot.  As we can see, the improvement is even better than in the case of the first-order derivative, and the error at all the boundaries has been reduced in two/three orders of magnitude.

\begin{figure}[t]
\begin{center}
\resizebox{.475\textwidth}{!}
{ 
\includegraphics{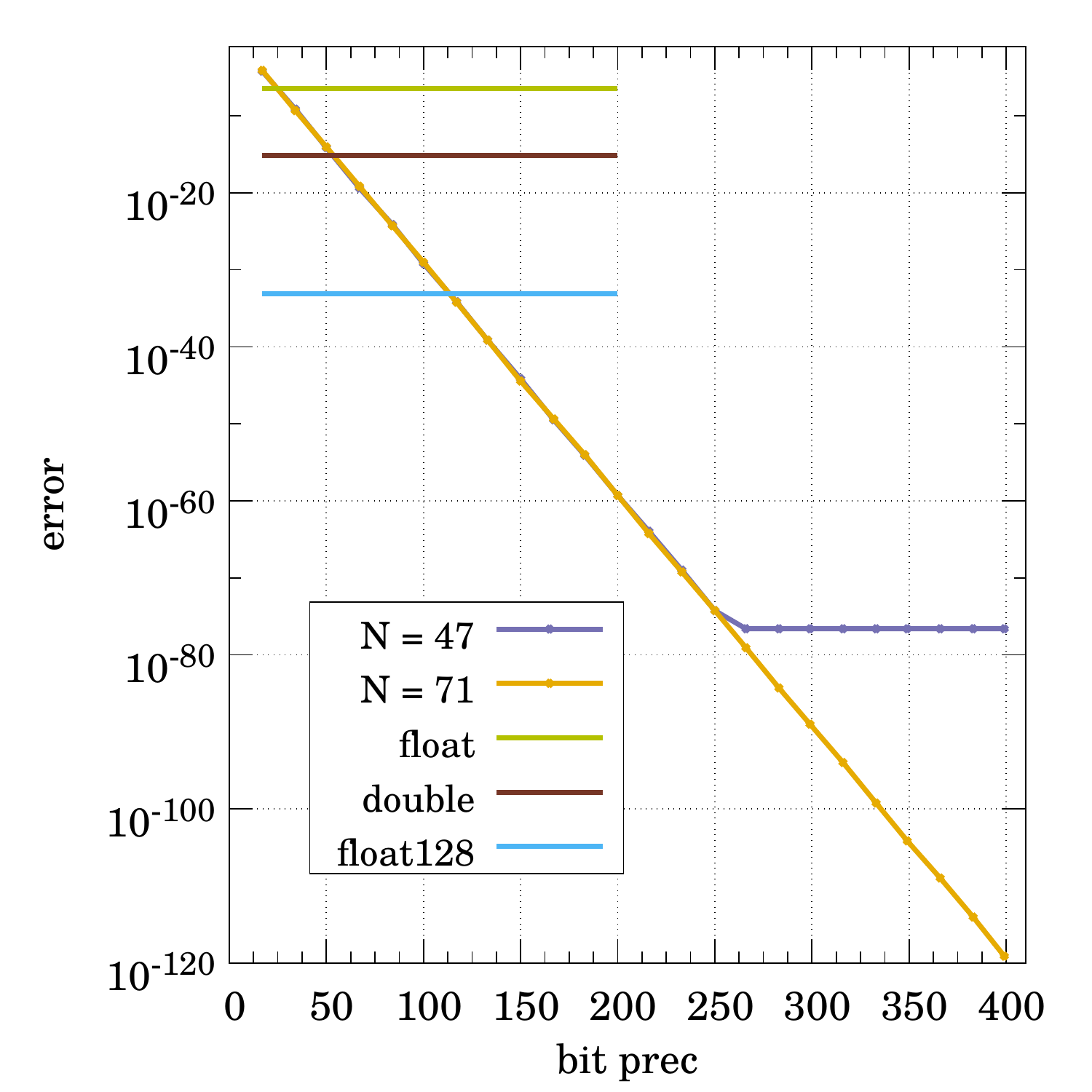}
}
\resizebox{.475\textwidth}{!}
{ 
\includegraphics{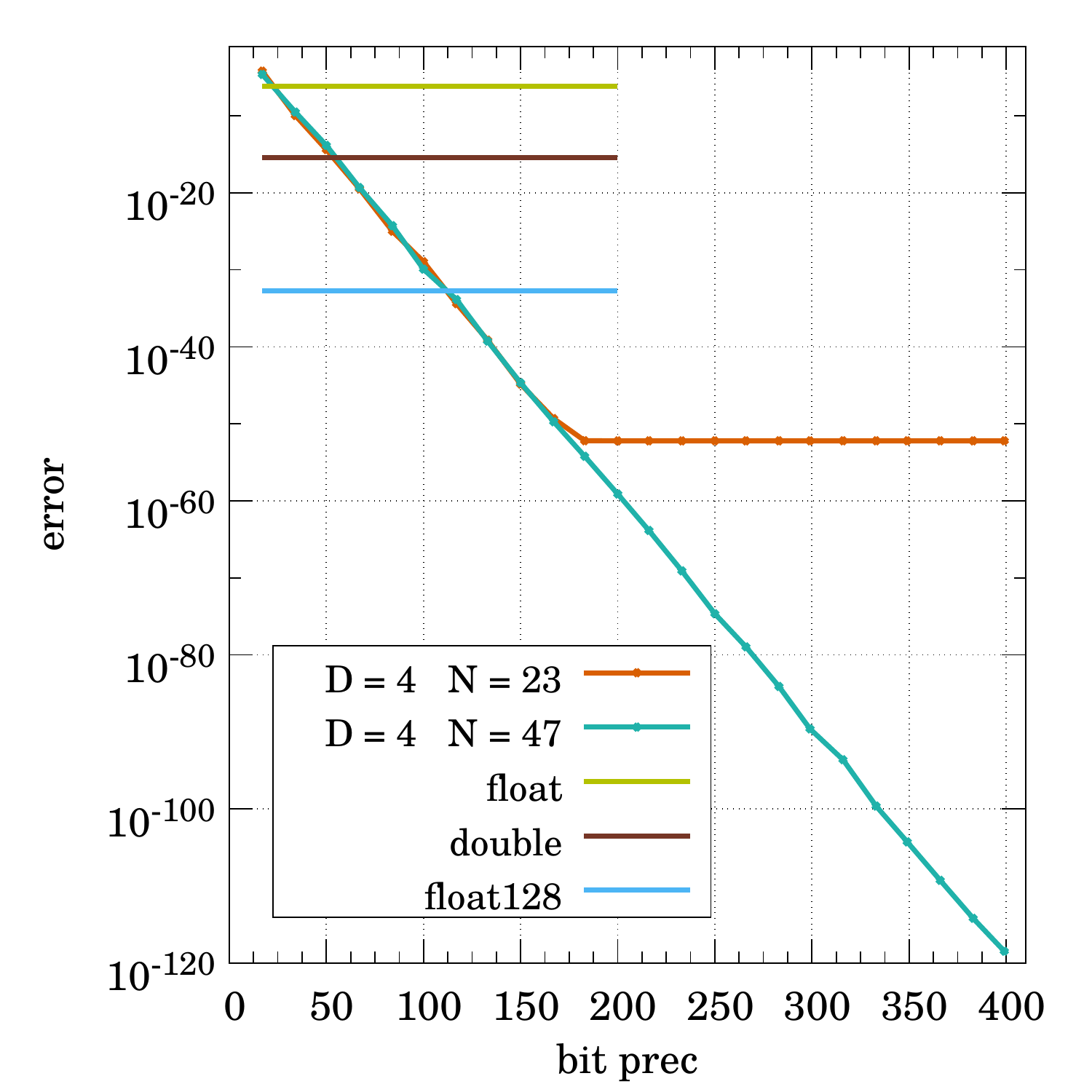}
}
\caption[Maximum Error for Integration and different Bit Precision.]{\textbf{Maximum Error for Integration and different Bit Precision.} The left plot shows the error made in the integration on a single Lobatto-Chebyshev collocation grid. The plot on the right shows the same error but when using the multidomain PSC method. In both cases, the points show the errors made for two different number of collocation points with respect to the bit precision used.  The horizontal lines indicate the error for single, double, and quadruple precision. \label{fig_err_integral} }
\end{center}
\end{figure}

Let us now consider the computation of the integral of the simple function $f(x) = \cos(x)\,$, for different bit precisions and both for single and multidomain setups. We show the results in Figure~\ref{fig_err_integral}. The plot on the left shows that as long as one uses enough collocation points the error scales with the round-off error as expected. This is true even for a number of collocation points as low as $N=47$ up to $260$ bits of precision, when the error saturates because the discretization error becomes more important than the round-off error. We also see that it is enough to use $N=71$ collocation points to decrease the error up to $10^{-120}$. The same behaviour is shown on the right plot of Figure~\ref{fig_err_integral} for the case in which we use a multidomain scheme with only four subdomains ($D=4$) and $N=23$ and $N=47$ collocation points per subdomain. In the first case, the discretisation error is reached around $10^{-50}$ but, for the second one we can easily reach again $10^{-120}$. The horizontal lines in the figure show the round-off error for the single, double, and quadruple precisions respectively, which is obviously much higher than the error we can reach using arbitrary precision. Although the integral under consideration involved a simple function, we have checked that the same is true independently of the integrand, just restricting ourselves to smooth functions.

\begin{figure}[t]
\begin{center}
\resizebox{0.95\textwidth}{!}
{ 
\includegraphics{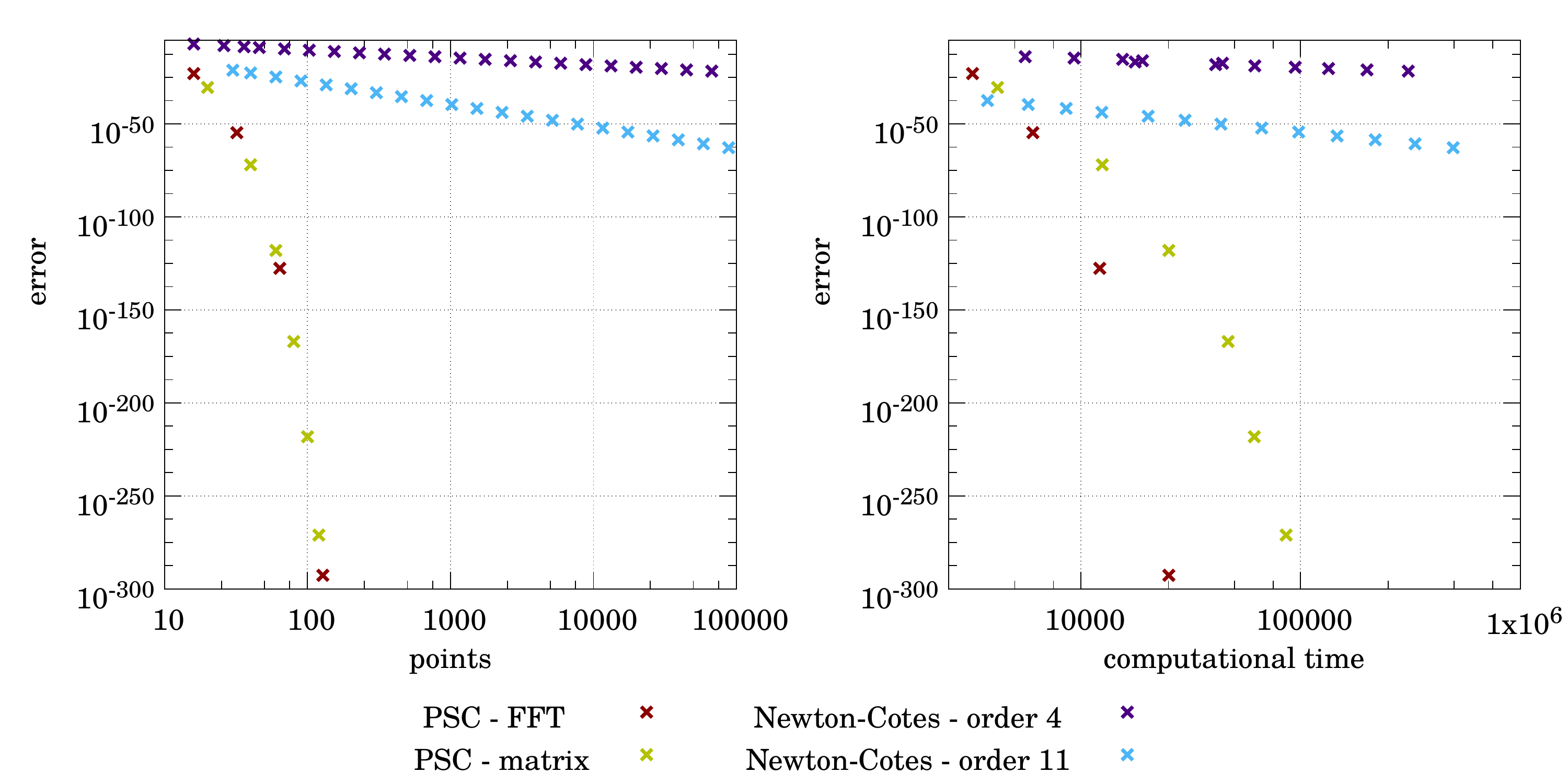}
}
\caption[Comparison between the PSC method and Newton-Cotes formulae for numerical integrals.]{\textbf{Comparison between the PSC method and Newton-Cotes formulae for numerical integrals.} We compare the error made in the computation of the integral of our test function using the PSC method, both with the matrix transformation (yellow points) and the FFT transformation (red), and two different Newton-Cotes formulae, one of fourth order (dark blue) and another one of eleventh order (light blue). In the left plot we compare the error made with the different methods in terms of the number of points used.  In the right plot we compare the computational time required by each method to achieve a certain level of accuracy.
\label{fig_err_integral_FD}}
\end{center}
\end{figure}

The accuracy that PSC methods are capable to reach is even more clear when we compare it with other discretization methods. In the left plot of Figure~\ref{fig_err_integral_FD}, we compare the results obtained using the ANETO library to Finite-Differences Newton-Cotes (NC) formulae~\cite{Press:1992nr} of fourth and eleventh orders. In both cases, we set manually the points near the boundaries of the subdomain to avoid the problems associated with the borders. The NC formula of eleventh order seems to be good enough to achieve a reasonable accuracy but the polynomial dependence of the error makes it very difficult to reach accuracies beyond $10^{-50}$, while with the PSC method accuracies of the order of $10^{-300}$ are within reach by using just $100$ collocation points.  Another relevant comparison is to look at the computational time required for a given accuracy for the two methods.  This is shown in the right plot of Figure~\ref{fig_err_integral_FD}. In the case of the PSC method, we distinguish the case of using the matrix transformation versus the FFT algorithm for the transformation between the physical and the spectral representations.  The plot clearly shows that the PSC method achieves high accuracies in much less time that the NC formulae.  It also shows that the use of the FFT algorithm for the transformation between representations is also more efficient than the matrix transformation as expected.  Going back to the comparison between the Finite Differences and PSC methods, it is clear that the development of a numerical code to implemented the NC formulae is always easier than the development of a PSC numerical code.  However, the PSC code does not need to be modified in order to go to very high precisions while improving the Finite Differences code for taking advantage of the potential accuracy usually requires to increase the order of the Finite Differences algorithm. This last option can be very challenging and, for high accuracy, it will not beat the exponential convergence of the PSC method.

\subsection{Double versus Arbitrary Precision: Computational Time}
\label{subsec_comp_time}

Most of the available computers nowadays use $64$-bit processors that are highly optimised to work with double precision.  Going beyond this precision usually requires a software implementation that slows down the computation.  The question is how much slow down can be expected depending on the precision that we need to use.

\begin{figure}[t]
\begin{center}
\resizebox{.85\textwidth}{!}
{ 
\includegraphics{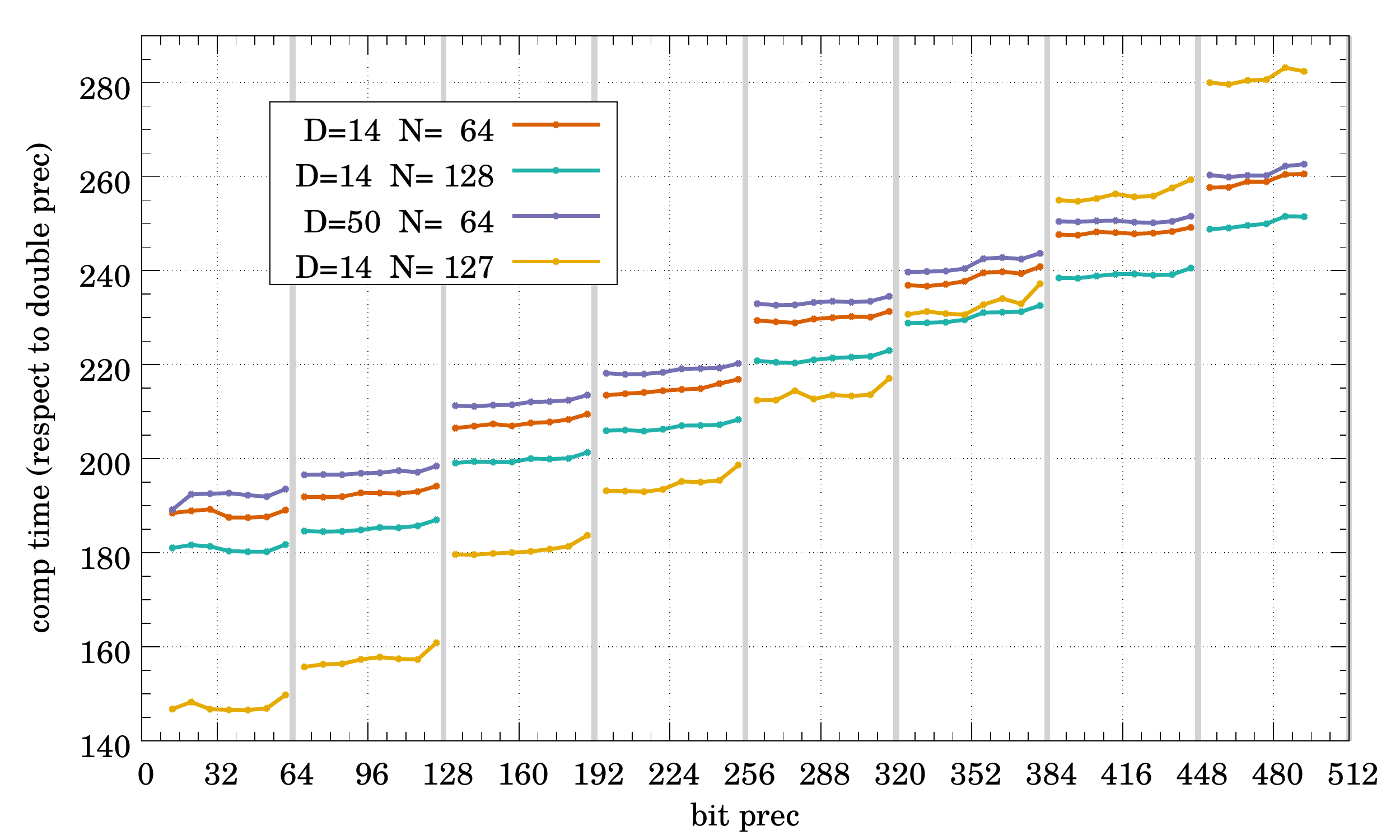}
}
\caption[Computational Time versus Arbitrary Precision]{\textbf{Computational Time versus Arbitrary Precision.} The plot shows the computational time employed in the integration of $f(x)=\cos(x)$ for four different grid configurations characterized by the number of subdomains, $D$, and the number of collocation points per subdomain, $N$. The computational times shown are relative to the computational time corresponding to double precision. The computational time appears to increase linearly with the bit precision. The vertical lines just separate multiples of $64$ bits. The computational time experiences a jump at these boundaries, as expected, due to the CPU architecture. The runs with $N=64$ and $N=128$ use the FFT algorithm to transform between the physical and spectral representations, while the one with $N=127$ uses the matrix transformation.  As a consequence, the computations for the grid configuration $(D,N)=(14,127)$ are much slower, in absolute computational time, than the ones for $(D,N)=(14,128)$.}
\label{fig_time}
\end{center}
\end{figure}

To answer this question we use the test case consisting in the integral again of the simple function: $f(x) = \cos(x)\,$.  We quantify the cost for four different grid configurations characterized by the pair $(D,N)$: (number of subdomains, number of collocation points per subdomain). The four grid configurations are: (i) $(D,N) = (14,64)$; (ii) $(D,N) = (14,128)$; (iii) $(D,N) = (50,64)$; and (iv) $(D,N) = (14,127)$. The results are presented in Figure~\ref{fig_time}, from where we can see how much slower is the use of arbitrary precision as compared with the standard double precision. In the range analysed, until $512$ bits, or around $150$ significant digits, the computational time seems to increase linearly with the number of bits, being a factor $150$-$300$ times slower than the double precision case.  Moreover, we have to take into account that in order to take advantage of the additional significant digits we need to increase the number of subdomains and/or collocation points.  Nevertheless, using the PSC method we do not need to change our numerical code, just the pair $(D,N)$.  These considerations are important to estimate whether arbitrary precision can be a good solution for a given numerical problem.

Looking at Figure~\ref{fig_time}, it is also worth mentioning the jumps in the computational cost every time we cross a vertical line (corresponding to multiples of $64$ bits). Although the general behaviour is linear, at small scales the function is more or less flat, increasing in multiples of $64$ bits. This is not surprising because the tests have been done using a $64$-bit processor and are related to the way the library uses double precision numbers to store the arbitrary floating points variables.

\subsection{Shared-memory parallelization of PSC computations using arbitrary precision}
\label{subsec_perf_openMP}

We have just seen that one of the drawbacks of using arbitrary precision is the loss of computational speed. Given that most current computer processors are designed for $64$-bit precision computations, arbitrary precision libraries usually emulate this using symbolic calculations or by means of a software layer that implements the arbitrary precision operations using $64$-bit data. Both options significantly slow down the computations that in our case can be of the order of $150-300$ times. 

\begin{figure}[t]
\begin{center}
\resizebox{.85\textwidth}{!}
{ 
\includegraphics{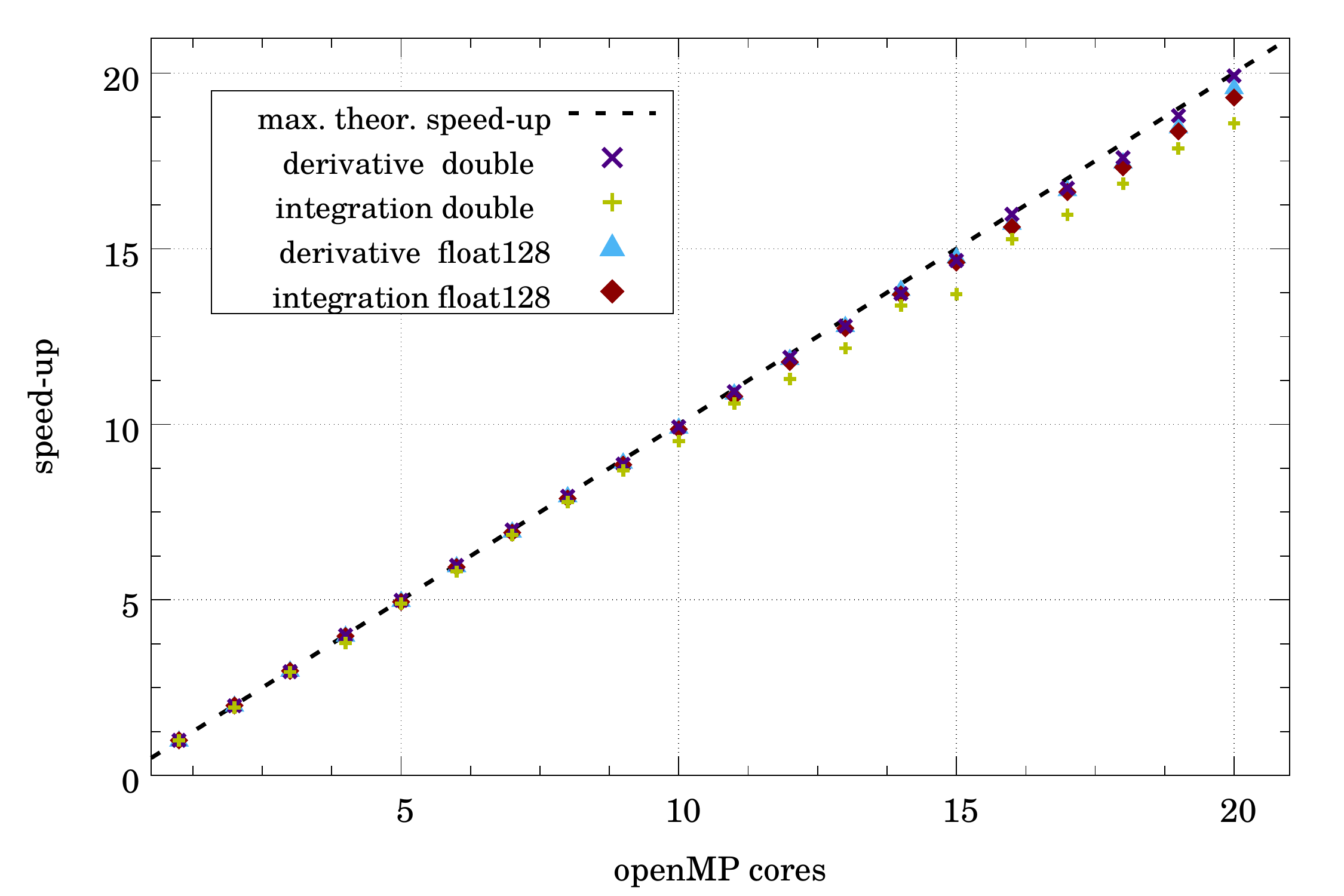}
}
\caption[Speed-up with OpenMP for multidomain PSC computations.]{\textbf{Speed-up with OpenMP for multidomain PSC computations.} This plot shows the computational speed-up, as defined in Eq.~(\ref{speedup-definition}), for integration and differentiation algorithms with both double and quadruple precision. Differentiation presents a better speed-up but both cases show that the multidomain method is a very good option for parallelization.}
\label{fig_openMP}
\end{center}
\end{figure}

In order to alleviate this downside of the method we can resort to parallelization.  Indeed, the use of parallel computing adapts perfectly to our PSC multidomain scheme because most of our computations are done independently within each of the subdomains. For instance, in the case of derivatives, they can be trivially parallelized because they are defined at each subdomain independently. Instead, in the case of global integrals (integrals over the whole computational domain) it is a bit more complex but they can be adapted without affecting the scaling with the number of operations. To see how we can parallelize integrals, let us look at the example of the indefinite integral of an arbitrary function $g(x)$ defined over the whole computational domain $\Omega$ and with a boundary condition at the left global boundary $x^{}_{L}$:     
\begin{equation}
I^{}_{\rm L}(x) = I^{b}_{\rm L} + \int^{x}_{x^{}_{L}} dx' g(x')\,.
\end{equation}
where $I^{b}_{\rm L} = I^{}_{\rm L}(x^{}_{L})$ is the boundary condition.  Since $x$ is an arbitrary point, it can belong to any subdomain $\Omega_{a}$ and then this integral may appear as a serial computation that is difficult to cast into a parallel one because the result depends on the integral for smaller values than $x^{}_{L}$.  Nevertheless, we can parallelize the computation in the following way: First of all, let us assume $x$ belongs to the subdomain $\Omega^{}_{a}$. Then, we can divide the integral into the sum of partial integrals that can be computed individually at each subdomain: 
\begin{equation}
I^{p}_{{\rm L},d} = \int^{x^{}_{d,R}}_{x^{}_{d,L}}dx\,g(x) = \int^{1}_{-1} dX\,\left(\frac{dx}{dX}\right)^{}_{X}g(x(X))  
\quad (d=0,\ldots,a-1)\,,
\label{partial_integral_0a-1}
\end{equation}
and the integral of the subdomain of $x$ ($\Omega^{}_{a}$):
\begin{equation}
I^{}_{{\rm L},a} (x) = \int^{x}_{x^{}_{a,L}}dx'\,g(x') = \int^{X(x)}_{-1} dX'\,\left(\frac{dx}{dX}\right)^{}_{X'}g(x(X'))\,,
\label{partial_integral}
\end{equation}
so that the full integral can be computed as:
\begin{equation}
I^{}_{\rm L}(x) = I^{b}_{\rm L} + \sum_{d=0}^{a-1} I^{p}_{{\rm L},d} + I^{}_{{\rm L},a}(x) \quad (x \in \Omega^{}_{a})\label{full_integrall}\,.
\end{equation}
With this separation, each piece in the sum can be computed independently in each subdomain. To implement this in practice, we propose the use of shared-memory parallelization, more specifically the broadly used application programming interface OpenMP~\cite{dagum1998openmp}.  This appears to be the simplest option to profit from the possibility of having independent computations in the different subdomains and adding the minimum possible communication overhead.  In order to test this we have carried out a number of numerical experiments using OpenMP.  The measure of the computational time speedup that we use, $S_p$, is a function of the number of cores employed, $p$, defined as:
\begin{equation}
S^{}_p = \frac{T}{T^{}_p} \,, 
\label{speedup-definition}
\end{equation}
where $T$ is the computational time spent by a sequential computation and $T^{}_p$ is the time corresponding to a computation that uses $p$ cores.  Of course, the ideal unreachable limit is $S^{}_p = p$.  Within this framework we have performed tests for integration and differentiation with the parallel multidomain PSC method.  This has been done both for double and quadruple ({\em float128}) precisions.  The results for the computational speedup $S_p$ are shown in Figure~\ref{fig_openMP}. We can see that we are very close to the maximum speed-up which is indicated by a dashed line in the plot. It is also interesting to note that the differentiation computations are closer to full parallelism than the integration ones.  This is expected from the fact that differentiation can be carried out fully independently at each subdomain while for integration we need to communicate the value of the partial integrals.

Up to here it is clear that the subdomain is the minimum unit of parallelization.  However, this approach does not address the principal reason for the slowdown, the emulation of the fundamental operations. This is a question currently under investigation in projects like~\cite{joldes:hal-01312858}, where arbitrary precision numbers are represented as an expansion of double precision numbers of different magnitudes and then the idea is to take advantage of parallel computations with graphics processing units (GPUs) to implement these basic operations.

\section{Gravitational Collapse in Asympotically-flat Spacetimes}\label{grav-collapse-Minkowski}

The first application of the numerical techniques we have just presented is the classical problem of gravitational collapse in a flat (Minkowski) spacetime (without cosmological constant, $\Lambda = 0$).  The energy-momentum distribution of the matter collapsing corresponds to a massless real scalar field. For simplicity we assume the spacetime to be spherically symmetric, which means that the Einstein field equations become $1+1$ PDEs (in time and in the radial direction).  Then, the setup of the gravitational dynamics to be followed numerically is quite simple: We consider initial states described by smooth initial data and such that the scalar field distribution is concentrated around a certain radial location. Then, there are only two possible end states for the evolution: \\

\noindent (i) {\em Collapse} of the scalar field and the formation of a Black Hole (BH).  \\

\noindent (ii) {\em Dispersion} of the scalar field with flat spacetime as the end state of the evolution.\\

To which one of these two states will the evolution drive the system depends on the features of the initial scalar field configuration, in particular on its energy density.  An interesting question is what separates these two very different outcomes of the evolution.  M. Choptuik~\cite{Choptuik:1992jv} carried out a systematic numerical study of this question and found that the dependence of the final state on the initial data is through a single (arbitrary) parameter. Moreover, Choptuik found that in the threshold between collapse and dispersion there is a one-parameter family of critical solutions that exhibit a naked singularity.  It was also found that the mass of the collapsed configurations near the threshold exhibits a scaling with a universal exponent.  These unexpected results attracted a lot of attention to this problem and constituted a cornerstone in the development of Numerical Relativity. A detailed review of critical gravitational collapse for different types of matter fields and spacetime configurations in General Relativity can be found in~\cite{Gundlach:2002sx,Gundlach:2007gc}.  

In this paper, like in the initial studies by Choptuik, we will restrict ourselves to problems in spherical symmetry.  We do not make any further simplification of the problem apart from this one.  The first step towards numerical simulations of gravitational collapse is to choose an adequate formulation of the Einstein field equations.  We start by choosing what is called a characteristic approach to the problem (introduced in~\cite{Christodoulou:1986zr,Goldwirth:1987nu,Garfinkle:1994jb}). The main idea of this approach is to set initial data on a null (or light-like) slide instead of a constant time slide as in a Cauchy-based initial-value problem.  The difference is that the normal to a null slide is a light-like one-form while the normal to a constant time slide is a space-like one-form, like in a standard initial-value Cauchy problem. The main advantage of the characteristic formulation is its ability to approach BH formation much more efficiently than a typical Cauchy one. It is worth mentioning that the initial Choptuik study~\cite{Choptuik:1992jv} was based on a Cauchy formulation and used adaptive mesh refinement to reach the necessary accuracy.  Later, Garfinkle~\cite{Garfinkle:1994jb} revisited the problem using a characteristic approach and recovered some of the main results without the need of refinement, which shows the power of the characteristic formulation for the study of gravitational collapse.  Our previous studies~\cite{Olivan:2015fmy,SantosOlivan:2016djn} confirmed the superior performance of the characteristic approach by evolving scalar fields, not in asymptotically-flat spacetimes but in Asymptotically-AdS (AAdS) spacetimes. However, the framework we set up in~\cite{Olivan:2015fmy,SantosOlivan:2016djn} is not suitable for the use of PSC methods, then we developed a Finite Differences numerical code. In this section we present a new and improved characteristic scheme that is adapted for the use of the PSC method.

The formulation of the characteristic problem goes as follows: Let us consider a self-gravitating massless scalar field, $\phi$.  The set of PDEs that we need to solve are the coupled system formed by the Einstein field equations
\begin{equation}
R^{}_{\mu\nu} - \frac{1}{2} g^{}_{\mu\nu} R + \Lambda g^{}_{\mu\nu}= 2 \, T^{}_{\mu\nu}\,.
\label{efes}
\end{equation}
and the equations for the scalar field which come from the energy-momentum conservation equations
\begin{equation}
\nabla^{\nu}\left(R^{}_{\mu\nu} - \frac{1}{2} g^{}_{\mu\nu} R + \Lambda g^{}_{\mu\nu} \right) = 0\quad \Longrightarrow \quad
\nabla^{\nu}T^{}_{\mu\nu} = 0 \,. \label{divT}
\end{equation}
In these equations, $R^{}_{\mu\nu}$ is the Ricci Tensor associated with the metric tensor $g^{}_{\mu\nu}$; $R$ is the scalar of curvature; $\Lambda$ is the cosmological constant; and $T^{}_{\mu\nu}$ is the energy-momentum tensor, which for a real massless scalar field is given by
\begin{equation}
T^{}_{\mu\nu} = \phi^{}_{;\mu} \phi^{}_{;\nu} - 2\;  g^{}_{\mu\nu} \phi^{}_{;\alpha}\phi^{;\alpha}\,,
\end{equation}
where the semicolon denotes covariant differentiation. Then, the resulting field equation for the scalar field is the well-known Klein-Gordon equation:
\begin{equation}
\Box \phi \equiv \phi_{;\mu}{}^{;\mu} =0\,.
\label{klein-gordon-equation}
\end{equation}

\begin{figure}[t]
\begin{center}
\resizebox{.75\textwidth}{!}
{ 
\includegraphics{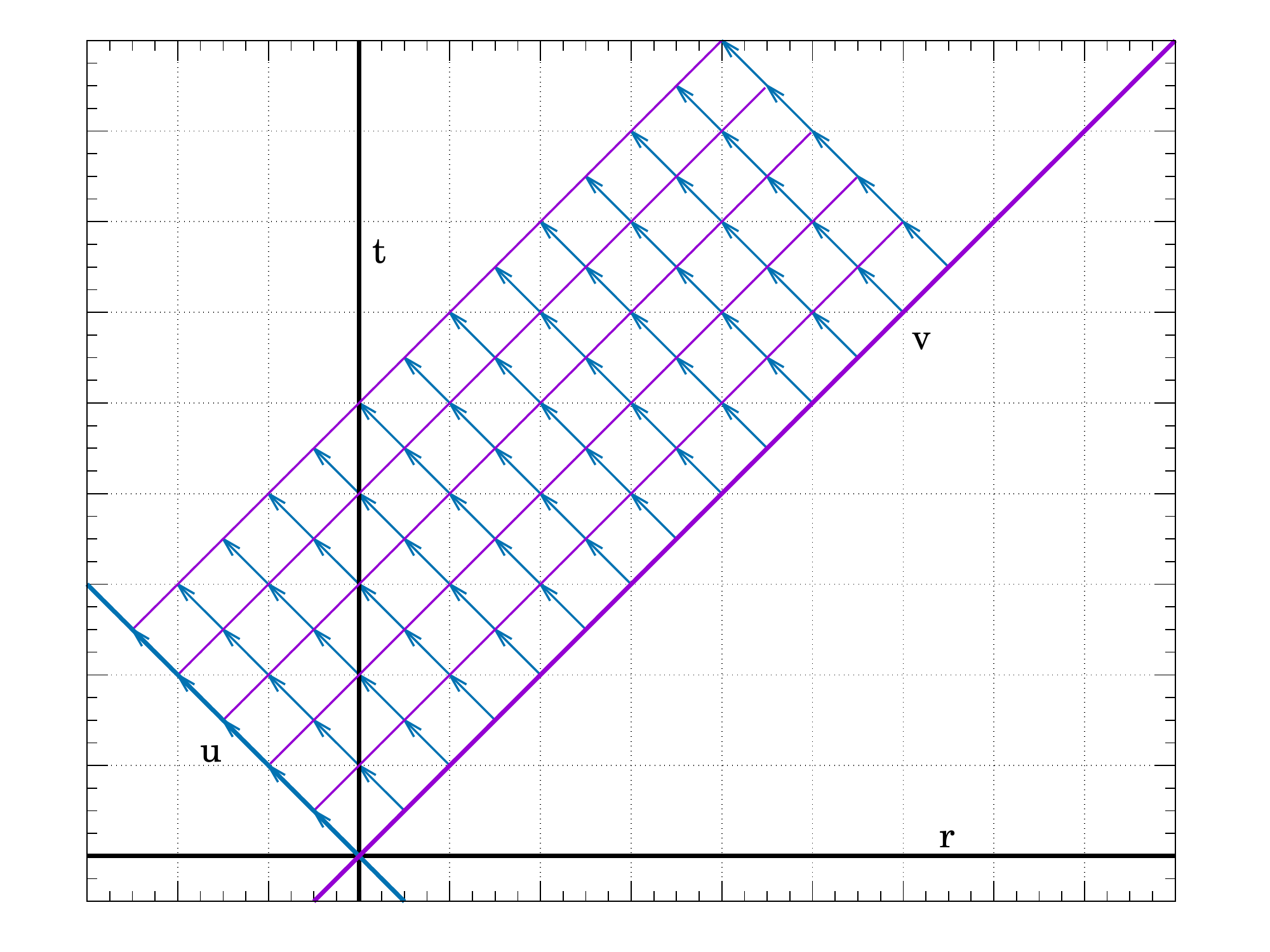}
}
\caption[Scheme of an Evolution using Double-Null Coordinates]{\textbf{Scheme of an Evolution using Double-Null Coordinates.} The horizontal and vertical lines are the axes corresponding to the time and radial coordinates, $(t,r)$. In a characteristic formulation, we set our initial conditions on a $u=$const slide (purple thick line) and evolve each point in the direction indicated by the arrows.
Here, constant null coordinate lines form $45$ degrees with respect to the axes $(t,r)$ but this is just a simplification of the drawing. \label{fig_doublenull_scheme}}
\end{center}
\end{figure}

To introduce the characteristic formulation we first need an adapted coordinate system.  We choose double-null coordinates for the time-radial section together with spherical coordinates for the spheres of symmetry of the problem.  The metric tensor in those coordinates is:
\begin{equation}
ds^2 = g^{}_{\mu\nu}dx^{\mu}dx^{\nu} = - 2 f(u,v)\, r^{}_v(u,v) \, dudv + r^2(u,v)\, d\Omega^2\,,
\label{line-element}
\end{equation}
where $(u,v)$ are the double-null coordinates ($\partial/\partial u$ and $\partial/\partial v$ are light-like vectors) and $d\Omega^{2}=d\theta^{2}+\sin^{2}\theta d\varphi^{2}$ is the line element of the unit 2-sphere. Moreover, $f$ and $r$ are two functions of $(u,v)$ and $r^{}_v$ is a shorthand for the partial derivative of $r$ with respect to the null coordinate $v$. In Figure~\ref{fig_doublenull_scheme} we show a representation of the characteristic grid for the case of an empty (flat) spacetime. In our case, the spacetime is curved by the presence of the massless scalar field $\phi$ and hence $(u,v)$ are not be perpendicular to the $(t,r)$ coordinate lines.

The set of PDEs for the components of the metric tensor, the metric functions $f(u,v)$ and $r(u,v)$, and the scalar field, $\phi(u,v)$, are obtained by introducing the metric in Eq.~(\ref{line-element}) into Eqs.~(\ref{efes}) and~(\ref{klein-gordon-equation}).  In order to reduce the order of the equations, from second-order to first-order PDEs, and to decouple them we introduce new variables associated with the scalar field $\phi$:
\begin{eqnarray}
h       & = & \frac{(r \phi)^{}_{v}}{r^{}_v}\,,\\
\bar{h} & = & \phi\,.
\end{eqnarray}
We also introduce a new metric variable:
\begin{equation}
\bar{f} = - 2 r^{}_u\,.
\label{def_fb}
\end{equation}
Then, the $vv$ and the $uv$ components of the Einstein field equations can be written as:
\begin{eqnarray}
f^{}_v        &=& \frac{f \, r^{}_v}{r} \left(h - \bar h\right)^2 \,, \\
\bar{f}^{}_{v} &=& \frac{r^{}_v}{r} \left( f - \bar f \right) \,.
\label{equation_fv_fbv}
\end{eqnarray}
In this characteristic formulation of the initial-value problem we prescribe initial conditions for the variable $h$ on an initial $u=u_{i}=$const. null slide, that is $h(u_{i},v)\,$.  We also need to prescribe $r(u_{i},v)$ and $r_{v}(u_{i},v)$.  With this information we can obtain the rest of variables, at the same null slide $u=u_{i}$, using the equations above.  The expressions for $h$, $f$, and $\bar{f}$ are:
\begin{eqnarray}
\bar{h}(u^{}_{i},v) &=& \frac{1}{r} \int_{v^{}_{o}(u^{}_{i})}^{v} d\tilde{v}\, h(u^{}_{i},\tilde{v}) \,  r^{}_v(u^{}_{i},\tilde{v})  \,,  
\label{char-eq-hbar} \\[1mm]
f(u^{}_{i},v)  &=& f(u^{}_{i},v^{}_{o}(u^{}_{i}))\,\exp\left\{ \int_{v^{}_{o}(u^{}_{i})}^{v} \!\!\!\! d\tilde{v}\, \frac{r^{}_v(u^{}_{i},\tilde{v})}{r(u^{}_{i},\tilde{v})} \left[h(u^{}_{i},\tilde{v})-\bar{h}(u^{}_{i},\tilde{v})\right]^2 \right\}\,, 
\label{char-eq-f} \\[1mm]
\bar{f}(u^{}_{i},v) &=& \frac{1}{r} \int_{v^{}_{o}(u^{}_{i})}^{v} d\tilde{v}\, f(u^{}_{i},v)\,r^{}_v(u^{}_{i},\tilde{v})   \,,
\label{char-eq-fbar}
\end{eqnarray}
where $v_o(u_{i})$ and $f(u^{}_{i},v^{}_{o}(u^{}_{i}))$ are the values of $v$ and $f(u,v)$ respectively, at the origin $r=0$ on the null slide $u=u_{i}$.   By looking at these expressions we realize that we need to guarantee the regularity of the different quantities at the origin, which translates into imposing the conditions: 
\begin{eqnarray}
\bar h(u^{}_{i},v^{}_{o}(u^{}_{i})) & = & h (u^{}_{i},v^{}_{o}(u^{}_{i}))\,, \\[1mm]
\bar f(u^{}_{i},v^{}_{o}(u^{}_{i})) & = & f (u^{}_{i},v^{}_{o}(u^{}_{i}))\,.
\end{eqnarray}
In this way, all the equations have a finite limit when we approach the origin $r=0$.  However, for numerical purposes it is not convenient to have divisions where both numerator and denominator approach zero.  This may be particularly problematic in the case of Eq.~(\ref{char-eq-fbar}), but we can transform it by using integration by parts and taking the right limits.   The result is:
\begin{equation}
\bar{f}(u^{}_{i},v) = f(u^{}_{i},v) - \frac{1}{r} \int_{v^{}_{o}(u^{}_{i})}^{v} \!\!\!\! d\tilde{v}\, f(u^{}_{i},\tilde{v}) r^{}_v(u^{}_{i},\tilde{v})  \left[ h(u^{}_{i},\tilde{v}) -\bar{h}(u^{}_{i},\tilde{v})\right]^{2} \,.
\end{equation}

Once we have $(h,\bar{h},f,\bar{f})$ at the null slide $u=u_{i}$ we can evolve them to the next null slide by using the evolution equation for the scalar field, i.e. Eq.~(\ref{klein-gordon-equation}), which comes from the energy-momentum conservation equation~(\ref{divT}). Due to the spherical symmetry only the scalar field has true dynamics since the degrees of freedom of the gravitational field are not activated in spherical symmetry.  The evolution equation to pass from a null slide to the next one is given by:
\begin{equation}
h^{}_{u}   =  \frac{1}{2\,r} \left(f - \bar f \right)  \left(h- \bar h \right)\,.
\label{dncharevol}
\end{equation}
This equation is actually an ODE for each value of $v$.  Indeed, let us consider a particular value of $v$, say $v_{\ast}$, then Eq.~(\ref{dncharevol}) takes the values of the variables at $(u_{i},v_{\ast})$ and gives us the value of $h$ at $(u_{i}+\Delta u,v_{\ast})$, being $\Delta u$ the time step used in the evolution.
In addition, from Eqs.~(\ref{def_fb}) and~(\ref{equation_fv_fbv}) we can obtain the evolution equations for $r$ and $r_{v}$:
\begin{eqnarray}
r^{}_u & = & - \frac{1}{2} \bar{f}\,, \label{eq_r_u} \\[1mm]
(r^{}_v)^{}_u & = & - \frac{1}{2} r^{}_v \left[ \frac{f - \bar{f}}{r} \right]\,. \label{eq_rv_u}
\end{eqnarray}
The only missing piece in this characteristic evolution scheme is the value of $f(u^{}_{i},v^{}_{o}(u^{}_{i}))$ that appears in Eq.~(\ref{char-eq-f}), which corresponds to the value of $f$ at the origin on the null slide $u=u_{i}$. This is a freely specifiable quantity that reflects the residual coordinate {\em gauge} freedom that we have in the choice of the null coordinate $v$.  This, in turn, can be seen as the remaining gauge freedom in completely specifying the radial function $r$ in our characteristic formulation.  In particular, it allows us to specify the location of the origin $r=0$ at the null slides $u=$const. Or in other words, the freedom in choosing the motion of the origin ($r=0$) as we evolve from one null slide to the next one. Indeed, since $r=r(u,v)$, and assuming that $v=v_{o}(u)$ corresponds to the location of the origin, i.e. $r(u,v_{o}(u))=0$, the equation of motion of the origin as we move through the spacetime foliation in null slides $u=$const. is given by:
\begin{equation}
\frac{dv^{}_{o}(u)}{du} = -\left.\frac{r^{}_{u}}{r^{}_{v}}\, \right|_{v=v^{}_{o}(u)} = \left. \frac{f^{}_{o}}{2\, r^{}_v}\, \right|^{}_{v=v^{}_{o}(u)}\,, \label{motion-of-origin}
\end{equation}
where we have used Eq.~(\ref{eq_r_u}).  We can then use this freedom to make the origin move, for instance, with a uniform speed.  To achieve this we just need to choose the freely specifiable quantity $f(u,v^{}_{o}(u))$  as: 
\begin{equation}
f(u,v^{}_{o}(u)) = 2 \left. r^{}_v \right|^{}_{v=v^{}_{o}(u)} \quad \Longrightarrow \quad v^{}_o(u) = v^{}_o(u^{}_i) + u\,.
\end{equation}

In our formulation the formation of an apparent horizon (AH) happens when the following condition is fulfilled: 
\begin{equation}
r^{}_v \longrightarrow 0\,, \quad \mbox{or equivalently,} \quad \frac{\bar{f}}{f} \longrightarrow 0\,.
\label{ah-formation-condition}
\end{equation}
This limit cannot be reached with our choice of system of coordinates (it corresponds to a coordinate singularity) although we can approach it as much as we want.  Then, we assume that an AH has formed when the quantities in the AH condition above reach a value less than $10^{-8}$. At that point we stop the simulation.

Finally, we used s Gaussian packets as initial data for the scalar field:
\begin{equation}
h(u^{}_i,v) =  \epsilon \; \exp\left\{ -\frac{(v - b)^{2}}{\omega^2} \right\}\,,
\label{initial-data}
\end{equation}
where the amplitude $\epsilon$, the width $\omega$, and the shift $b$ are the freely specifiable parameters of this 3-parameter family of initial data.  

At this point we have presented all the necessary ingredients for the  characteristic formulation of the problem. It is well-adapted for its implementation using the PSC method for the discretization in the $v$ coordinate. We evolve from one null slide $u=$const. to the next one by using a standard Runge-Kutta 4 (RK4) algorithm. We have implemented this formulation using the PSC method and arbitrary precision tools described in the previous sections.  The error can be estimated from the absolute value of the last spectral coefficient. The value of the coefficients, $a_n$ ($n=0,\ldots,N$), decay exponentially, reaching or not round-off error. If round-off is not reached, the last spectral coefficient represents an estimation of the truncation error incurred in ignoring the rest of terms in the spectral series. Otherwise, the last coefficient represents the precision reached. In both cases, it can be used for a good estimation of the error. 

\begin{figure}[t]
\begin{center}
\resizebox{.48\textwidth}{!}
{ 
\includegraphics{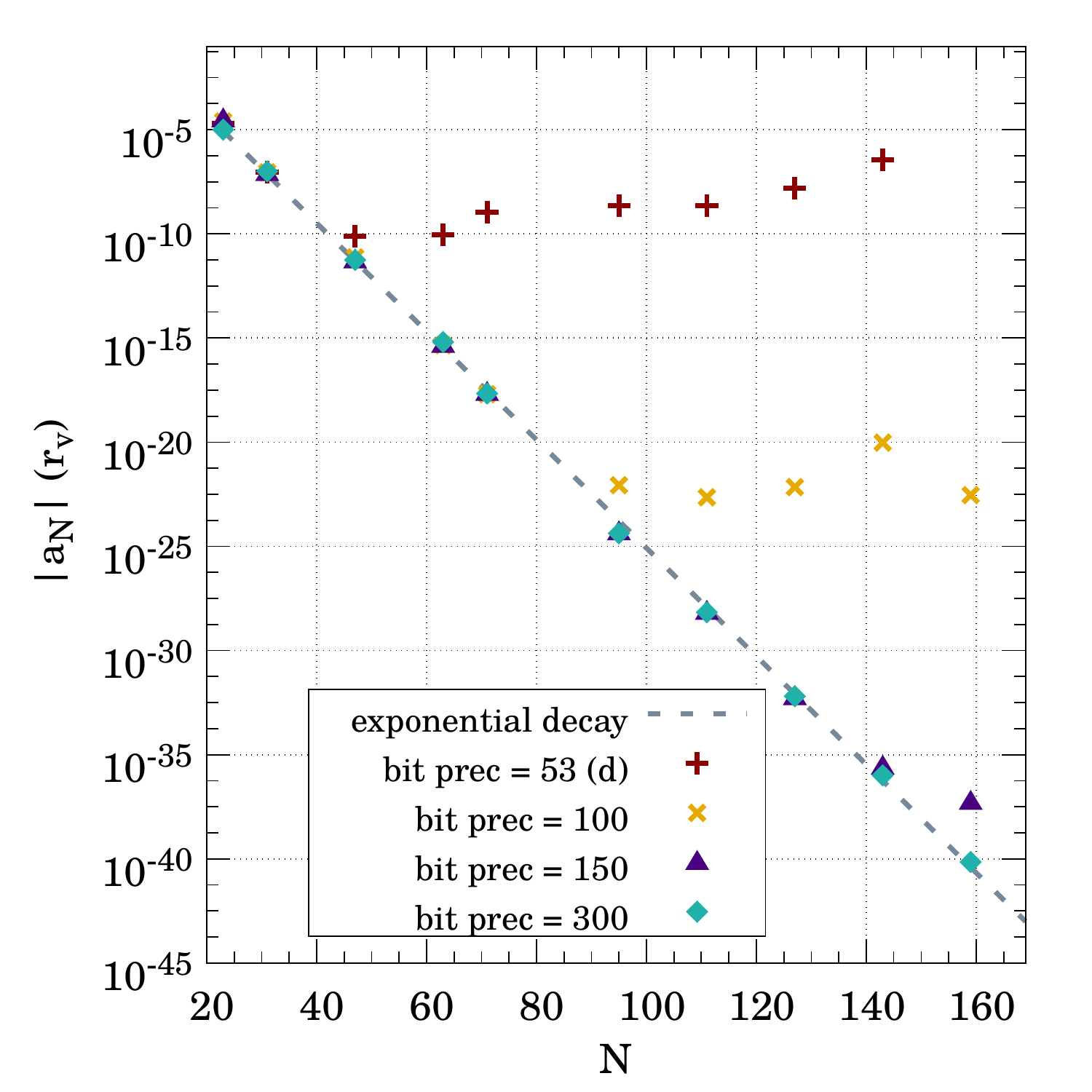}
}
\resizebox{.48\textwidth}{!}
{ 
\includegraphics{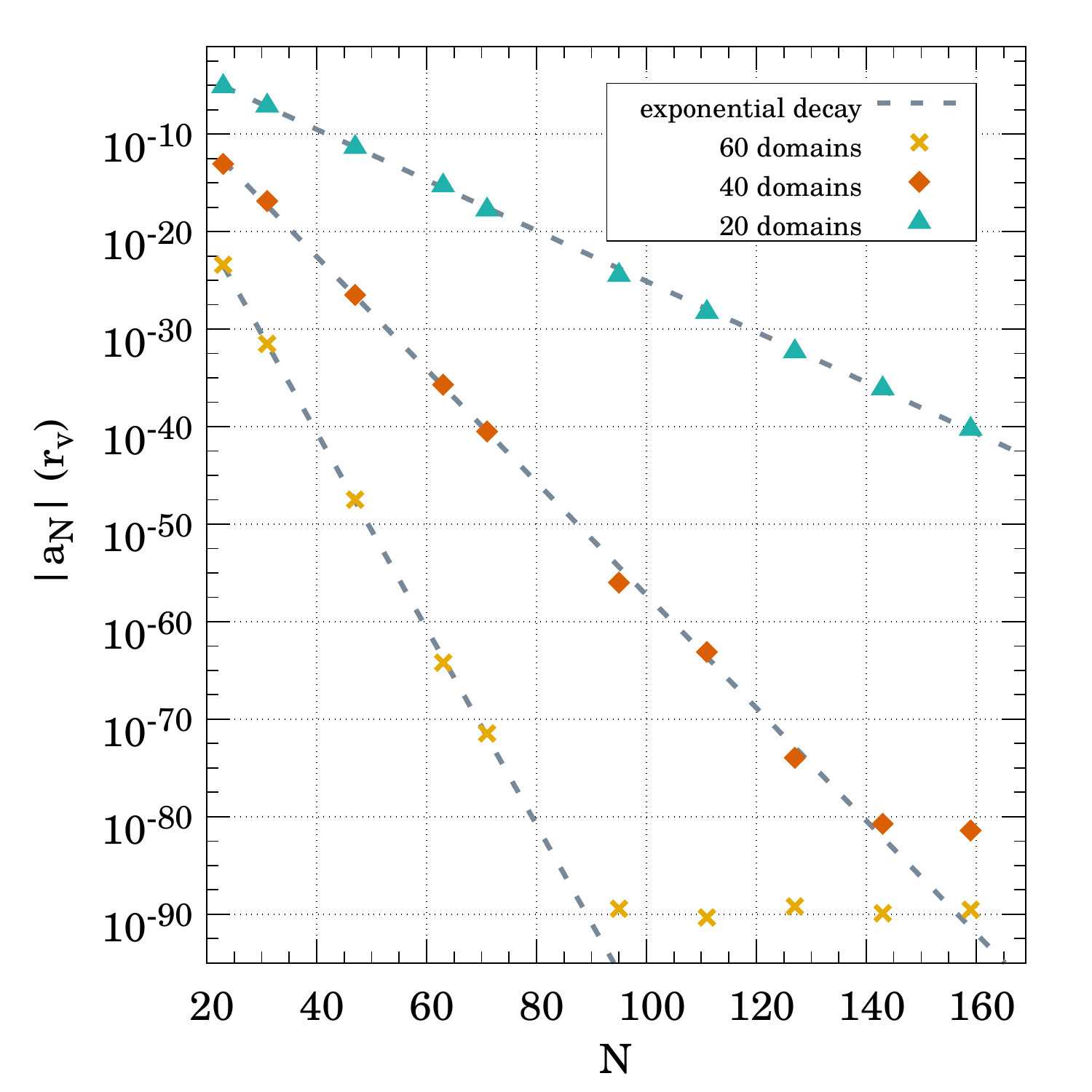}
}
\caption[Convergence in the estimation of the location of the formation of an Apparent Horizon (AH)]{\textbf{Convergence in the estimation of the location of the formation of an Apparent Horizon (AH):} The left plot shows the truncation error (estimated from the last spectral coefficient) at the moment of AH formation for several different grid configurations, all of them with $D=20$ subdomains. Each data set corresponds to simulations done with different bit precision. The error decays exponentially (spectral convergence) until the maximum precision is reached. On the right plot we show the impact of adding more subdomains. All data sets exhibit spectral convergence but the number of subdomains has an impact in the $\alpha$ factor of the exponential decay $e^{-\alpha N}$. These simulations use $300$ bit precision.
\label{fig_gravColl_error}}
\end{center}
\end{figure}

We have evolved the initial data of Eq.~(\ref{initial-data}) with parameters: $\epsilon = 2.00$, $b= 0.15$ and $\omega = 0.05$.  This is an example of initial configuration above the critical threshold so that it will collapse and form an AH.  In Figure~\ref{fig_gravColl_error} we show the error in the location of the AH at the end of the evolution in terms of the number of collocation points per subdomain.  Since we have different errors at each subdomain, we take the highest of all of them, which corresponds to the subdomain where the collapse takes place. In Figure~\ref{fig_gravColl_error} we only show the error associated with the function $r_v$, to monitor the first condition in Eq.~(\ref{ah-formation-condition}), because is the one that has the highest error of all of the three evolution variables.

Moreover, on the left plot of Figure~\ref{fig_gravColl_error}, we use a setup with $D=20$ subdomains and change the number of collocation points for different bit precisions. In all the cases, the error has an exponential decay (spectral convergence) until we rearch the precision limit (round-off error) which, of course, improves as we increase the number of bits of our data types.   The first one represents a $53$ bit precision, equivalent to the standard double precision, which allows us to obtain a maximum accuracy of $10^{{\rm -}10}$-$10^{{\rm -}11}$. Notice that this is few orders of magnitude above the theoretical limit of sixteen digits.  This fact is not surprising since we have to consider that during the evolution the numerical noise piles up, reducing the maximum precision. In addition, the number of subdomains used in our test evolutions is not optimal. This is just a comparison of the same exact setup for several bit precisions. Increasing the number of significant bits we improve the maximum error and with $150$ bits (around $45$ significant digits) we easily decrease the error up to almost $10^{-40}$.  On the right plot of Figure~\ref{fig_gravColl_error} we study the influence of the number of subdomains in the error as we change the number of collocation points per subdomain but keeping the number of subdomains constant.  The error presents an exponential decay $|\Delta r_v| \approx \exp(-\alpha N)$. Varying the number of subdomains changes the factor of the exponential decay, $\alpha$. In this case we can reach the minimum error by adding subdomains with less collocation points. This can be a good idea considering that adding subdomains has, in general, a linear impact on the computational time while increasing the number of collocation points, $N$, has an impact of $\sim N \log N$ or $\sim N^2$ depending on whether the operations in the spectral domain are performed using a FFT transformation or a matrix transformation respectively.

\section{(In)Stability of Anti-de Sitter Spacetimes}\label{grav-collapse-anti-de-Sitter}

In this section we consider a new physical scenario for the application our hybrid PSC-arbitrary precision method.  We present results of the evolution of (exact non-linear) ``perturbations'' in asymptotically Anti-de Sitter spacetimes using a Cauchy formulation of the initial-value problem.  Cauchy-type evolutions in spherical symmetry were already done in the study of critical gravitational collapse by Choptuik~\cite{Choptuik:1992jv}.  In the context of Anti-de Sitter spacetimes they were used recently to study also the problem of critical gravitational collapse in~\cite{Bizon:2011gg}.  We adapted this formulation for the use of the PSC method in Refs.~\cite{Olivan:2015fmy,SantosOlivan:2016djn} and we found new physical features associated with the non-linear evolution.  We now present results from a new adaptation of our numerical scheme to include arbitrary precision together with the PSC method.

Anti-de Sitter spacetimes have attracted a lot of attention in the last years, both for the interest in studying the non-linear (in)stability of AdS and for its relevance in the so-called AdS/CFT correspondence (also known as the gauge/gravity duality). The key feature of AAdS spacetimes is the presence of a boundary that light-like signals (light rays, massless fields, etc.) can reach in a finite time, but such that time-like signals (massive particles, massive fields, etc.) will take infinite proper time to reach.  This property changes completely the landscape of gravitational collapse.  As a consequence, the two-case scenario of asymptotically-flat spacetimes does not apply to AAdS spacetimes.  Indeed, considering an initial profile like in Eq.~(\ref{initial-data}), we can also expect that the dynamics will make this configuration either to collapse [case (i)] or to disperse [case (ii)].  In the case the scalar field configuration collapses it will form a BH that eventually will settle down into a stationary state (a Schwarzschild-Anti-de Sitter BH). However, in the case the scalar field disperses we cannot expect this dispersion to proceed until we reach asymptotically AdS.  The scalar field propagates locally at the speed of light and then, after some finite time, the scalar field profile will reach the AdS boundary, it will bounce back and will try to collapse again, only that the non-linear evolution will change the profile and we will be in the initial situation but with different initial conditions.  Therefore, there will be again two possible outcomes, collapse to form an AH or dispersion until reaching the AdS boundary in a finite time and bounce back. This process will repeat itself until the scalar field configuration will have a profile dense enough to finally collapse forming an AH.  Then, the possible states of the evolution of a scalar profile in spherically-symmetric AAdS spacetimes are: \\

\noindent (1) {\em Direct Collapse} of the scalar field and the formation of a Black Hole (BH).  \\

\noindent (2) {\em Collapse} of the scalar field and BH formation after $1$ bounce off the AdS boundary. \\

$\vdots$\\

\noindent ($n_{c}$) {\em Collapse} of the scalar field and BH formation after $n_{c}-1$ bounces off the AdS boundary. \\

During the trip to the AdS boundary and back, the non-linear relativistic evolution induces a transfer of energy from low frequency (long wavelength) modes towards high frequency (short wavelength) modes, similar to what happens in the onset of turbulence.  Due to this analogy, this process, that at some point will end in the collapse of the scalar field profile, has been named the {\it turbulent instability} of AdS spacetime.  That is, no matter how small would be the amplitude of the initial profile (perturbation), the field will finally collapse and the end state would be an AdS-BH spacetime. Nevertheless, although this turbulent instability may appear to be a generic feature of the dynamics in AdS spacetime, there are indications of the existence of some islands of stability (see, e.g.~\cite{Choptuik:2018ptp} and references therein) that will not follow the channels just described.  The reason is that some stable configurations have been found for some forms of initial configurations, but the exact extend of these ``stability islands'' in the parameter space of initial configurations is still under debate. In order to study this question we need extremely long and accurate evolutions. In conclusion, the end state of the evolution of perturbations in AdS spacetimes is an ideal testbed for numerical techniques that provide high accuracy, beyond the standard one, as the problem is highly demanding.  In order to illustrate this, in this section we present a test case in which we evolve an initial massless scalar field configuration in AAdS spacetimes during two of these bounces and compare the accuracy using double precision with the accuracy using $300$-bit precision.

In order to solve the Einstein field equations [Eq.~(\ref{efes})] coupled to the massless scalar field equation [Eq.~(\ref{klein-gordon-equation})] in this new scenario we need a different coordinate system.  This coordinate system has to be adapted to a Cauchy-type initial-value problem and, at the same time, it has to incorporate the AdS asymptotic structure of the spacetime.  With this in mind, the form of the spacetime line element that we consider is~\cite{Bizon:2011gg}:
\begin{equation}
ds^2 = g^{}_{\mu\nu}dx^{\mu}dx^{\nu} = \frac{\ell^{2}}{\cos^{2}x}\left( - A {e}^{-2\delta}\,dt^2 +\frac{dx^{2}}{A} + \sin^2 x\, d\Omega^2 \right)\,, \label{aads_metric}
\end{equation}
where $A = A(t,x)$ and $\delta = \delta(t,x)$, $t$ is the time coordinate, and $x$ is a compactified radial coordinate in such a way that the AdS boundary is located at $x = \pi/2$ instead of at infinity.  The overall factor contains $\ell$, the AdS length scale, which is related to the negative cosmological constant of the spacetime, $\Lambda<0$ [see Eq.~(\ref{efes})], by the expression: $\ell^2 = -3/\Lambda$.  The time coordinate $t$ has an infinite range, i.e. $t\in$ ($-\infty$,$\infty$), whereas $x$, being compactified, goes from $x=0$ (center) to $\pi/2$ (AdS boundary). We can recover AdS spacetime by setting $A = 1$ and $\delta=0$.

The system of PDEs that be obtain, by choosing the right combination of variables, can be reduced to a first-order system of strongly hyperbolic PDEs. In addition, in order to use our multidomain scheme, we can further specialize our variables and take them to be the characteristic variables of the hyperbolic system~\cite{Courant:1989aa,SantosOlivan:2016djn}. The form of the characteristic variables associated with the scalar field that we adopt is:
\begin{eqnarray}
U & = & \frac{1}{\cos x} \left( \phi^{}_x - \frac{e^{\delta}}{A} \phi^{}_t  \right) \,, 
\label{Udef} \\
V & = & \frac{1}{\cos x} \left( \phi^{}_x + \frac{e^{\delta}}{A}\phi^{}_t  \right) \,, 
\label{Vdef}
\end{eqnarray}
where again, the $(t,x)$ subscripts denote partial differentiation with respect to these coordinates. Then, using the $(t,x)$ coordinates and the $(U,V)$ variables, the evolution problem is reduced to the following coupled system of PDEs: 
\begin{eqnarray}
U^{}_t & = & - A e^{-\delta} U^{}_{x} - \frac{(3-2 \cos^2 x)}{\sin x \cos x}  U \,e^{-\delta}\,(1-A) 
         - \frac{A e^{-\delta} }{\sin x \cos x} \left( U+V \right)  \nonumber \\
       & + & \frac{\sin x}{\cos x}U\,Ae^{-\delta} \,,  \label{U-dot-eq}         \\[1mm]
V^{}_t & = & + A e^{-\delta} V^{}_{x} + \frac{(3-2 \cos^2 x)}{\sin x \cos x}  V \,e^{-\delta}\,(1-A) 
         + \frac{A e^{-\delta} }{\sin x \cos x} \left( U+V \right) \nonumber \\
       & - & \frac{\sin x}{\cos x}V\,Ae^{-\delta} \,.  
\label{V-dot-eq} 
\end{eqnarray}
It is also convenient to introduce the following normalized variable associated with the scalar field:
\begin{equation}
\psi = \frac{\phi}{\cos^{2}x} \,. \label{psi-def}
\end{equation}
This new scalar field variable satisfies both an evolution equation
\begin{eqnarray}
\psi^{}_t & = & \frac{A e^{-\delta}}{2\cos x} \left( V - U\right)  \,,  
\label{evol_psi}
\end{eqnarray}
and also a constraint equation (only containing spatial derivatives):
\begin{eqnarray}
\psi^{}_x =  2\,\frac{\sin x}{\cos x}\,\psi  + \frac{1}{2}\, \frac{U + V}{\cos x} \,.
\label{psi_prime}
\end{eqnarray}
Then, we can solve for $\psi$ either by evolving Eq.~(\ref{evol_psi}) or by solving this constraint equation on a constant time slide.  Regarding the metric functions, we do not expect them to satisfy hyperbolic equations since in spherical symmetry the true gravitational degrees of freedom are turned off.  Then, we obtain constraint equations for $\delta$ and $A$:
\begin{eqnarray}
\delta^{}_x & = & - \onehalf \sin x \cos^{3} x  \left( V^2 + U^2 \right)\,,
\label{delta_prime} \\[1mm]
A^{}_x  & = &  \frac{1+2\sin^2 x}{\sin x\cos x}(1-A) -  \frac{A}{2}\sin x\cos^{3}x\left( V^2 + U^2\right) \,,   
\label{A_prime} 
\end{eqnarray}
from which $\delta(t,x)$ and $A(t,x)$ can be obtained at a given time, once we have the solution for $(U,V)$ via the evolution equations, by performing the following integrals:
\begin{eqnarray}
\delta(t,x) & = & \int_x^{\frac{\pi}{2}}\!\! dy\, \sin y \cos^{3} y \left( \frac{U^2 + V^2}{2} \right)\,,  
\label{eq-delta-cons}\\[1mm]
A(t,x)  - 1 & = & -\frac{\cos^3 x\;e^{\delta}}{\sin x} \int_0^{x}\!\! dy\, e^{-\delta}\sin^{2}y \left( \frac{U^2 + V^2}{2} \right) \,,   
\label{ads_integration_for_A}  
\end{eqnarray}
where the boundary conditions are $A = 1$ both at $x=0$ and $x=\pi/2$ and we have chosen $\delta \left(\pi/2 \right) = 0$, fixing the time coordinate $t$ as the proper time at the AdS boundary.  Then, the Cauchy evolution goes as follows: (i) We prescribe initial data on an initial Cauchy surface $t=t_{o}=$const.  for $(U,V)$, i.e. $U(t_o, x)$ and $V(t_o, x)$. (ii) Using equations~(\ref{psi_prime}),~(\ref{eq-delta-cons}), and~(\ref{ads_integration_for_A}) we find $\psi(t_{o},x)$, $\delta(t_{o},x)$ and $A(t_{o},x)$. (iii) With this information we evolve $(U,V)$ from $t_{o}$ to $t_{o}+\Delta t$ using the evolution equations~(\ref{U-dot-eq}) and~(\ref{V-dot-eq}) and the boundary conditions.  In our numerical simulations we have considered the following family of Cauchy initial data:
\begin{eqnarray}
U (t^{}_o, x) = \epsilon \exp \left\{ - \frac{ 4 \tan^2 x }{\pi^2 \sigma^{2}} \right\}\,, 
\quad
V (t^{}_o, x) = - U (t^{}_{o}, x)\,,
\end{eqnarray}
where the freely specifiable parameters $\epsilon$ and $\sigma$ (amplitude and width of the Gaussian profile respectively) are chosen to be: $\epsilon = 2.0$ and $\sigma = 0.4$.  

We set up a multidomain grid with $D=10$ subdomains and change the number of collocation points per subdomain to see how the error changes for both double precision and for $300$-bit precision. We evolve the initial ``perturbation'' for the time corresponding to two bounces off the AdS boundary. Since we need very high accuracy, we have used a sixth-order Runge-Kutta 10,6(7) (see Refs.~\cite{verner1978explicit,prince1981high} for details). This ODE solver uses ten intermediate steps to generate a sixth-order accurate in time integration with a seventh-order step that is used as an estimation of the error.

\begin{figure}[t]
\begin{center}
\resizebox{.9\textwidth}{!}
{ 
\includegraphics{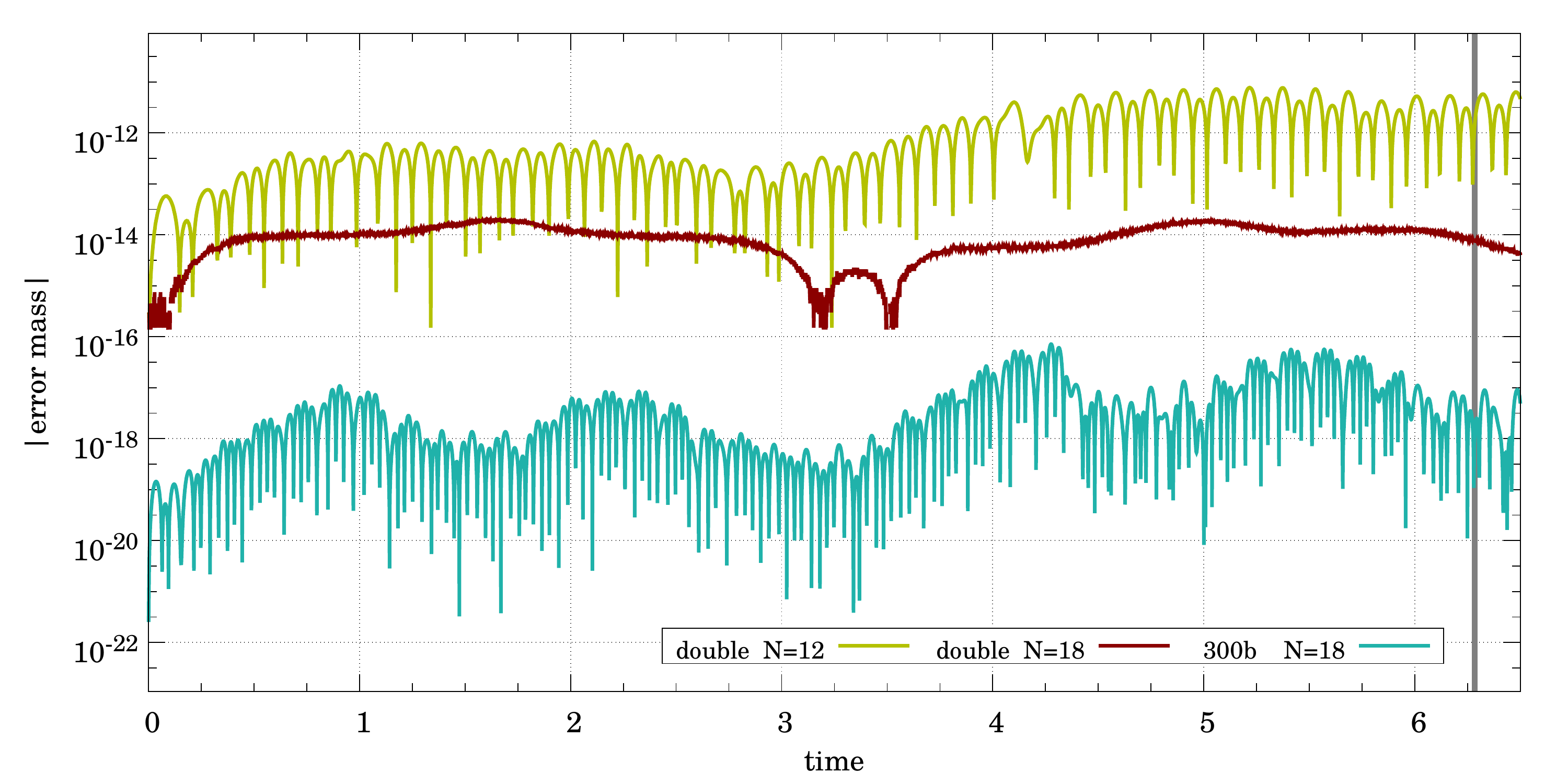}
}
\caption[Evolution of the Mass Error Function in AAdS Spacetimes]{\textbf{Evolution of the Mass Error Function in AAdS Spacetimes.} We compare three different grid/precision configurations with $D=10$ subdomains. Using double precision, the use of $N=18$ collocation points per subdomain (red line) is enough to reach the round-off error. The same number of collocation points but with $300$-bit precision (turquoise line) allows us to reduce a few orders of magnitude the error during the evolution. It is interesting to notice how, when the error is determined by the discretization error, several fluctuations are present while in the case dominated by machine round-off the error remains almost flat during the time evolution. The grey vertical line to the right shows the instant of time at which we measure the error for the study presented in Figure~\ref{fig_AdS_error}.
\label{fig_AdS_evol}}
\end{center}
\end{figure}

At any time step, we can compute the energy contained inside a sphere of a given compactified radius $x$, which is known as the mass function:
\begin{equation}
\mathcal{M}(t,x)  =  e^{\delta} \int_0^{x}\!\! dy\,  e^{-\delta}  \sin^{2} y  \left( \frac{U^2 + V^2}{2}\right) \,,
\end{equation}
so that the total energy contained in the spacetime is $M(t) = \mathcal{M}(t,\pi/2)$.  One can show that $M(t)$ should not depend on time, i.e. it is a conserved quantity.  Then, we can monitor the conservation of $M$ as indicator of the accuracy of the simulation. To that end we introduce the following mass error function:
\begin{equation}
\Delta{M}(t) = \frac{| M(t) - M(t^{}_{o}) |}{M(t^{}_{o})}\,.
\label{mass-error-function}
\end{equation}
We have performed a set of numerical evolutions to assess the relevance of arbitrary precision in these computations.  In Figure~\ref{fig_AdS_evol} we show the evolution of the mass error function of Eq.~(\ref{mass-error-function}). The purple line shows an evolution with a low number of collocation points ($N=12$) and with double precision. The error oscillates in the range $10^{\m 12}-10^{\m 11}$. Increasing the number of collocation points to $N=18$ (red line in Figure~\ref{fig_AdS_evol}) we can decrease the mass error down to around $10^{\m 14}$. It is interesting to notice the different behaviour between these two lines. In the first case, there is plenty of oscillations and we estimate the error by taking the maximum value. This is due to the fact that here the error is determined by the discretization error in such a way that the total error can oscillate between the discretization and the round-off errors. In the second case we have reached by far the round-off error and the profile is quite flat. Then, it is obvious that it cannot be improved by using double precision. Just changing from double precision to $300$-bit precision but with the same number of collocation points per subdomain, the error drops more than two orders of magnitude and, again, it is determined by the discretization.

\begin{figure}[t]
\begin{center}
\resizebox{\textwidth}{!}
{
\includegraphics{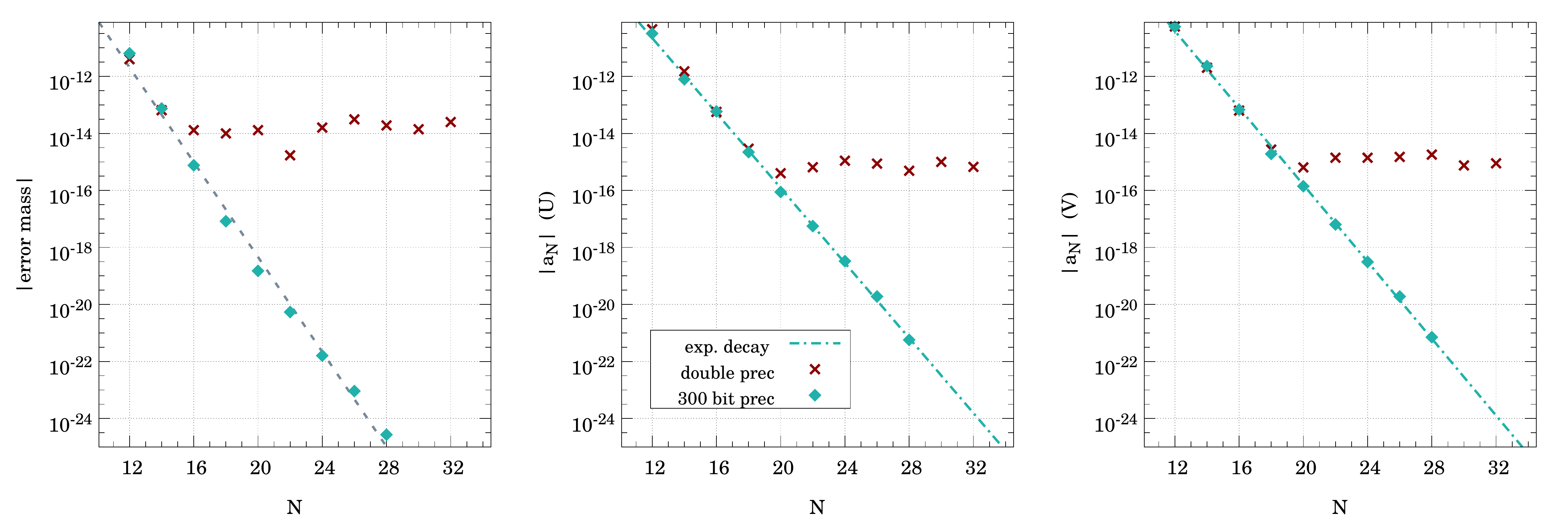}
}
\caption[Convergence of the Truncation Error in Evolutions in AdS Spacetimes]{\textbf{Convergence of the Truncation Error in Evolutions in AdS Spacetimes.} The figure shows the error in an AdS spacetime evolution using double (red dots) and 300-bit (turquoise dots) precision computations. The plots represent the normalised mass error $\Delta{M}$ (left), the truncation error for $U$ (centre), and the truncation error for $V$ (right) with respect to the number of collocation points per subdomain $N$. This information is taken at the same time ($t=2\pi$) after the massless scalar field configuration has  bounced twice off the AdS boundary. All the simulations use $D=10$ subdomains.
\label{fig_AdS_error}}
\end{center}
\end{figure}

Once the massless scalar field configuration has bounced twice off the AdS boundary and has come back to the initial location ($t \approx 2\pi$), we have studied not only the error mass function at that moment, but also the error in the characteristic fields $U$ and $V$.  To that end, we have used the absolute value of the last spectral coefficient in the subdomain where the error is maximum. This is shown in Figure~\ref{fig_AdS_error} for configurations with $D=10$ subdomains and for different number of collocation points and numerical precisions. As expected, the error in the three quantities [$\Delta{M}$ (left plot), $U$ (center plot), and $V$ (right plot)] decays exponentially until we reach round-off error. For double precision (red points) this happens at values of the order of $\sim 10^{\m 14}-10^{\m 15}$. This is easily improved when we use higher-order precision, as in the case shown with turquoise dots in the same figure.  This corresponds to $300$-bit precision and allows us to evolve the scalar field profile with an accuracy below $10^{\m24}$ with a number of collocation points per subdomain as small as $N=28$, for a total of $(N+1)\, D = 290$ collocation points.

\section{Conclusions and Future Perspectives}
\label{conclusions-perspectives}

In this paper we have shown the potential of the combination of Pseudo-Spectral Collocation methods and arbitrary-precision arithmetic for the solution of ordinary/partial differential equations, and more specifically for hyperbolic problems related to the description of gravitational collapse in relativistic gravitation. The exponential convergence of the PSC method makes it a very suitable choice for reaching the maximum accuracy associated with a certain bit precision with a relatively low number of discretization (collocation) points as compared with other techniques. In this sense, we have seen that the power-law convergence of finite difference algorithms makes it unfeasible to reach the needed accuracy within a reasonable number of discretization points. In addition, the PSC method does not require relevant changes in the algorithms as we increase the number of precision bits, in contrast with finite difference algorithms, where we need to adapt the algorithm so that the error scales in a way that we can reach the level of accuracy allowed by the choice of precision arithmetic.  

In Sec.~\ref{subsec_comp_time}, we have seen that the main problem of arbitrary precision arithmetic is that it is usually implemented via a software layer that slows down significantly the computations with respect to the speed of standard double precision arithmetic. In this sense, the PSC method helps since the number of collocation points required is relatively small and therefore, although the sparsity of the matrices involved in certain algorithms can be a drawback. Moreover, the multidomain scheme proposed in this paper allows for a simple parallelization of the computations, as we have shown with the use of OpenMP in our examples.  We have also shown that the scalability of the multidomain scheme is close to the ideal case of full parallelism.   Nevertheless, it would be desirable to explore improvements in the computation speed (and cost) based on an exploration of a more low-level approach to arbitrary precision arithmetic.

To illustrate the potential of these methods we have shown simulations in two problems in relativistic gravitational collapse: (i) The classical Choptuik collapse.  Here we have seen that we can estimate with arbitrary precision the location of the apparent horizon. (ii) Collapse in asympotically anti-de Sitter spacetimes.  In this example we have shown that arbitrary precision arithmetic allows us to preserve the total energy along the numerical evolution to a very high degree of precision. These numerical experiments have been carried out using a new library, the ANETO library~\cite{anetolib}, that we have developed in the course of our numerical studies of gravitational collapse in General Relativity.  The current version has been released with a few basics tools to deal with evolution problems but it can be extended in the future to include other tools that the PSC methods offers.  In this sense, one of the main possible improvements would be to add a solver for linear ODEs and also the incorporation of tools for non-linear systems. In addition, it would be interesting to add some type of Adapting Mesh Refinement to allow the grid to be more flexible under different conditions.  At the moment, the library, like the systems analyzed in this work, can only deal with evolution problems in just one spatial dimension.  It would be desirable to change in a future in order to incorporate tools to work on higher dimensional problems. Another aspect that we have not discussed much in this paper is the question of time integration. As the demand for accuracy increases, this aspect becomes more and more important and then, high-order integration algorithms would be required both in relation to the accuracy provided by the PSC method and to the one provided by arbitrary-precision arithmetic, otherwise we may be in a situation in which the evolution takes a considerably large number of time steps. An interesting solution to improve the accuracy of the evolutions with arbitrary precision arithmetic could be a spectral time integration like the one proposed in Refs.~\cite{Hennig:2008af,Hennig:2012zx,Macedo:2014bfa}.

\section*{Acknowledgements}
The authors acknowledge the high-performance computing resources provided by the Consorci de Serveis Universitaris de Catalunya (CSUC) and the Galicia Supercomputing Center (CESGA) under projects ICTS-CESGA-249 and ICTS-CESGA-266. They also acknowledge support from contracts ESP2015-67234-P, and ESP2017-90084-P (Spanish Ministry of Economy and Competitivity, MINECO), and from contract 2017-SGR-1469 from AGAUR (Catalan government). DS acknowledges support from a FPI doctoral contract BES-2012-057909 from MINECO. We also acknowledge networking support by the COST Action GWverse CA16104 (Horizon 2020 Framework Programme of the European Union).


\begin{thebibliography}{10}
\expandafter\ifx\csname url\endcsname\relax
  \def\url#1{{\tt #1}}\fi
\expandafter\ifx\csname urlprefix\endcsname\relax\def\urlprefix{URL }\fi
\providecommand{\eprint}[2][]{\url{#2}}

\bibitem{Lehner:2001wq}
Lehner L 2001 {\em Class. Quant. Grav.\/} {\bf 18} R25--R86 (\textit{Preprint}
  \eprint{gr-qc/0106072})

\bibitem{Grandclement:2007sb}
Grandclement P and Novak J 2009 {\em Living Rev. Rel.\/} {\bf 12} 1
  (\textit{Preprint} \eprint{0706.2286})

\bibitem{Alcubierre:1138167}
Alcubierre M 2008 {\em {Introduction to 3+1 Numerical Relativity}\/}
  International series of monographs on physics (Oxford: Oxford Univ. Press)
  \urlprefix\url{https://cds.cern.ch/record/1138167}

\bibitem{Bona:2009bo}
Bona C, Palenzuela-Luque C and Bona-Casas C 2009 {\em {Elements of Numerical
  Relativity and Relativistic Hydrodynamics: From Einstein's equations to
  Astrophysical Simulations}\/} (Berlin: Springer)

\bibitem{Baumgarte:2010bs}
Baumgarte T~W and Shapiro S~L 2010 {\em {Numerical Relativity: Solving
  Einstein's Equations on the Computer}\/} (Cambridge: Cambridge University
  Press)

\bibitem{Lehner:2014asa}
Lehner L and Pretorius F 2014 {\em Ann. Rev. Astron. Astrophys.\/} {\bf 52}
  661--694 (\textit{Preprint} \eprint{1405.4840})

\bibitem{Cardoso:2014uka}
Cardoso V, Gualtieri L, Herdeiro C and Sperhake U 2015 {\em Living Rev.
  Relativity\/} {\bf 18} 1 (\textit{Preprint} \eprint{1409.0014})

\bibitem{math3020337}
Bailey D~H and Borwein J~M 2015 {\em Mathematics\/} {\bf 3} 337 ISSN 2227-7390
  \urlprefix\url{http://www.mdpi.com/2227-7390/3/2/337}

\bibitem{Khanna:2013}
Khanna G 2013 {\em J. Sci. Comput.\/} {\bf 56} 366--380 ISSN 0885-7474
  \urlprefix\url{http://dx.doi.org/10.1007/s10915-012-9679-3}

\bibitem{canuto2007spectral}
Canuto C, Hussaini M~Y, Quarteroni A and Zang T~A 2007 {\em {Spectral Methods:
  Evolution to Complex Geometries and Applications to Fluid Dynamics}\/}
  ({Springer Science \& Business Media})

\bibitem{Bourke1988}
Bourke W 1988 {\em {Spectral Methods in Global Climate and Weather Prediction
  Models}\/} (Dordrecht: Springer Netherlands) pp 169--220 ISBN
  978-94-009-3041-4
  \urlprefix\url{http://dx.doi.org/10.1007/978-94-009-3041-4_4}

\bibitem{ehrendorfer2012spectral}
Ehrendorfer M 2012 {\em {Spectral Numerical Weather Prediction Models}\/}
  {Other titles in Applied Mathematics} (Society for Industrial and Applied
  Mathematics) ISBN 9781611971989

\bibitem{Szilagyi:2009qz}
Szilagyi B, Lindblom L and Scheel M~A 2009 {\em Phys. Rev.\/} {\bf D80} 124010
  (\textit{Preprint} \eprint{0909.3557})

\bibitem{Haas:2016cop}
Haas R {\em et~al.\/} 2016 {\em Phys. Rev.\/} {\bf D93} 124062
  (\textit{Preprint} \eprint{1604.00782})

\bibitem{Canizares:2009ay}
Canizares P and Sopuerta C~F 2009 {\em Phys. Rev.\/} {\bf D79} 084020
  (\textit{Preprint} \eprint{0903.0505})

\bibitem{Canizares:2010yx}
Canizares P, Sopuerta C~F and Jaramillo J~L 2010 {\em Phys. Rev.\/} {\bf D82}
  044023 (\textit{Preprint} \eprint{1006.3201})

\bibitem{Macedo:2014bfa}
Panosso R and Ansorg M 2014 {\em J. Comput. Phys.\/} {\bf 276} 357--379
  (\textit{Preprint} \eprint{1402.7343})

\bibitem{Oltean:2018szc}
Oltean M, Sopuerta C~F and Spallicci A~D~A~M 2018 {\em {Journal of Scientific
  Computing}\/} ISSN 1573-7691 (\textit{Preprint} \eprint{1802.03405})
  \urlprefix\url{https://doi.org/10.1007/s10915-018-0873-9}

\bibitem{anetolib}
{D Santos-Oliv\'an and C F Sopuerta} 2017 {ANETO library: Arbitrary precisioN
  solvEr with pseudo-specTral MethOds}
  \url{https://github.com/DSantosO/anetolib}

\bibitem{Olivan:2015fmy}
Santos-Oliv\'an D and Sopuerta C~F 2016 {\em Phys. Rev. Lett.\/} {\bf 116}
  041101 (\textit{Preprint} \eprint{1511.04344})

\bibitem{SantosOlivan:2016djn}
Santos-Oliv\'an D and Sopuerta C~F 2016 {\em Phys. Rev.\/} {\bf D93} 104002
  (\textit{Preprint} \eprint{1603.03613})

\bibitem{Choptuik:1992jv}
Choptuik M~W 1993 {\em Phys. Rev. Lett.\/} {\bf 70} 9--12

\bibitem{Hawking:1973uf}
Hawking S~W and Ellis G~F~R 1973 {\em {The Large Scale Structure of
  Space-Time}\/} (Cambridge: Cambridge University Press)

\bibitem{ieee745-2019}
{Institute of Electrical and Electronics Engineers (IEEE)} 2019 {IEEE 754-2019
  - IEEE Standard for Floating-Point Arithmetic}
  \url{https://standards.ieee.org/content/ieee-standards/en/standard/754-2019.html}

\bibitem{Boyd}
Boyd J~P {2001} {\em {Chebyshev and Fourier Spectral Methods}\/} 2nd ed (New
  York: Dover)

\bibitem{Fornberg:1996psc}
Fornberg B 1996 {\em {A Practical Guide to Pseudospectral Methods}\/}
  (Cambridge: Cambridge University Press)

\bibitem{Canutoetal:2006sm1}
Canuto C, Hussaini M~Y, Quarteroni A~M and Jr Z~T 2006 {\em {Spectral Methods.
  Fundamentals in Single Domains}\/} (Berlin Heidelberg: Springer-Verlag)

\bibitem{boostmultiprecisionlib}
{J Maddock and C Kormanyos} 2002 {Boost Multiprecision Library}
  \url{http://www.boost.org/doc/libs/1_66_0/libs/multiprecision/doc/html/index.html}

\bibitem{Fousse:2007}
Fousse L, Hanrot G, Lef\`{e}vre V, P{\'e}lissier P and Zimmermann P 2007 {\em
  ACM Trans. Math. Softw.\/} {\bf 33} ISSN 0098-3500
  \urlprefix\url{http://doi.acm.org/10.1145/1236463.1236468}

\bibitem{mpfrcpp}
Holoborodko P 2008-2012 {MPFR C++} http://www.holoborodko.com/pavel/mpfr/

\bibitem{dagum1998openmp}
Dagum L and Menon R 1998 {\em {Computational Science \& Engineering, IEEE}\/}
  {\bf 5} 46--55

\bibitem{canuto1988spectral}
Canuto C, Hussaini M~Y, Quarteroni A and Zang T~A 1988 {\em {Spectral Methods
  in Fluid Dynamics}\/} (Springer-Verlag)

\bibitem{Courant:1989aa}
Courant R and Hilbert D 1989 {\em {Methods of Mathematical Physics Volume
  II}\/} (John Wiley and Sons)

\bibitem{John:1991fj}
John F 1991 {\em {Partial Differential Equations}\/} (New York: Springer
  Verlag)

\bibitem{Sopuerta:2005gz}
Sopuerta C~F and Laguna P 2006 {\em Phys. Rev.\/} {\bf D73} 044028
  (\textit{Preprint} \eprint{gr-qc/0512028})

\bibitem{Canizares:2011kw}
Canizares P and Sopuerta C~F 2011 {\em Class. Quant. Grav.\/} {\bf 28} 134011
  (\textit{Preprint} \eprint{1101.2526})

\bibitem{Runge:1901cfr}
Runge C 1901 {\em {Zeitschrift f\"ur Mathematik und Physik}\/} {\bf 46}
  224--243

\bibitem{Epperson:1987je}
Epperson J 1987 {\em Amer. Math. Monthly\/} {\bf 94} 329--341

\bibitem{Press:1992nr}
Press W~H, Flannery B~P, Teukolsky S~A and Vetterling W~T 1992 {\em {Numerical
  Recipes: The Art of Scientific Computing}\/} (Cambridge (UK) and New York:
  Cambridge University Press)

\bibitem{joldes:hal-01312858}
Joldes M, Muller J~M, Popescu V and Tucker W 2016 {{CAMPARY: Cuda Multiple
  Precision Arithmetic Library and Applications}} {\em {5th International
  Congress on Mathematical Software (ICMS)}\/} (Berlin, Germany)
  \urlprefix\url{https://hal.archives-ouvertes.fr/hal-01312858}

\bibitem{Gundlach:2002sx}
Gundlach C 2003 {\em Phys. Rept.\/} {\bf 376} 339--405 (\textit{Preprint}
  \eprint{gr-qc/0210101})

\bibitem{Gundlach:2007gc}
Gundlach C and Martin-Garcia J~M 2007 {\em Living Rev. Rel.\/} {\bf 10} 5
  (\textit{Preprint} \eprint{0711.4620})

\bibitem{Christodoulou:1986zr}
Christodoulou D 1986 {\em Commun. Math. Phys.\/} {\bf 105} 337--361

\bibitem{Goldwirth:1987nu}
Goldwirth D~S and Piran T 1987 {\em Phys. Rev.\/} {\bf D36} 3575

\bibitem{Garfinkle:1994jb}
Garfinkle D 1995 {\em Phys. Rev.\/} {\bf D51} 5558--5561 (\textit{Preprint}
  \eprint{gr-qc/9412008})

\bibitem{Bizon:2011gg}
Bizo\'n P and Rostworowski A 2011 {\em Phys. Rev. Lett.\/} {\bf 107} 031102
  (\textit{Preprint} \eprint{1104.3702})

\bibitem{Choptuik:2018ptp}
Choptuik M, Santos J~E and Way B 2018 {\em Phys. Rev. Lett.\/} {\bf 121} 021103
  (\textit{Preprint} \eprint{1803.02830})

\bibitem{verner1978explicit}
Verner J~H 1978 {\em SIAM Journal on Numerical Analysis\/} {\bf 15} 772--790

\bibitem{prince1981high}
Prince P~J and Dormand J~R 1981 {\em {Journal of Computational and Applied
  Mathematics}\/} {\bf 7} 67--75

\bibitem{Hennig:2008af}
Hennig J and Ansorg M 2009 {\em J. Hyperbol. Diff. Equat.\/} {\bf 6} 161
  (\textit{Preprint} \eprint{0801.1455})

\bibitem{Hennig:2012zx}
Hennig J 2013 {\em J. Comput. Phys.\/} {\bf 235} 322--333 (\textit{Preprint}
  \eprint{1204.4220})

\end{thebibliography}

\providecommand{\newblock}{}

\end{document}